\documentstyle[11pt,amssymbols]{article}
\newread\epsffilein    
\newif\ifepsffileok    
\newif\ifepsfbbfound   
\newif\ifepsfverbose   
\newdimen\epsfxsize    
\newdimen\epsfysize    
\newdimen\epsftsize    
\newdimen\epsfrsize    
\newdimen\epsftmp      
\newdimen\pspoints     
\def\epsfbox#1{\global\def\epsfllx{72}\global\def\epsflly{72}%
   \global\def\epsfurx{540}\global\def\epsfury{720}%
   \def\lbracket{[}\def\testit{#1}\ifx\testit\lbracket
   \let\next=\epsfgetlitbb\else\let\next=\epsfnormal\fi\next{#1}}%
\def\epsfgetlitbb#1#2 #3 #4 #5]#6{\epsfgrab #2 #3 #4 #5 .\\%
   \epsfsetgraph{#6}}%
\def\epsfnormal#1{\epsfgetbb{#1}\epsfsetgraph{#1}}%
\def\epsfgetbb#1{%
\openin\epsffilein=#1
\ifeof\epsffilein\errmessage{I couldn't open #1, will ignore it}\else
   {\epsffileoktrue \chardef\other=12
    \def\do##1{\catcode`##1=\other}\dospecials \catcode`\ =10
    \loop
       \read\epsffilein to \epsffileline
       \ifeof\epsffilein\epsffileokfalse\else
          \expandafter\epsfaux\epsffileline:. \\%
       \fi
   \ifepsffileok\repeat
   \ifepsfbbfound\else
    \ifepsfverbose\message{No bounding box comment in #1; using defaults}\fi\fi
   }\closein\epsffilein\fi}%
\def\epsfclipstring{}
\def\epsfsetgraph#1{%
   \epsfrsize=\epsfury\pspoints
   \advance\epsfrsize by-\epsflly\pspoints
   \epsftsize=\epsfurx\pspoints
   \advance\epsftsize by-\epsfllx\pspoints
   \epsfxsize\epsfsize\epsftsize\epsfrsize
   \ifnum\epsfxsize=0 \ifnum\epsfysize=0
      \epsfxsize=\epsftsize \epsfysize=\epsfrsize
      \epsfrsize=0pt
     \else\epsftmp=\epsftsize \divide\epsftmp\epsfrsize
       \epsfxsize=\epsfysize \multiply\epsfxsize\epsftmp
       \multiply\epsftmp\epsfrsize \advance\epsftsize-\epsftmp
       \epsftmp=\epsfysize
       \loop \advance\epsftsize\epsftsize \divide\epsftmp 2
       \ifnum\epsftmp>0
          \ifnum\epsftsize<\epsfrsize\else
             \advance\epsftsize-\epsfrsize \advance\epsfxsize\epsftmp \fi
       \repeat
       \epsfrsize=0pt
     \fi
   \else \ifnum\epsfysize=0
     \epsftmp=\epsfrsize \divide\epsftmp\epsftsize
     \epsfysize=\epsfxsize \multiply\epsfysize\epsftmp
     \multiply\epsftmp\epsftsize \advance\epsfrsize-\epsftmp
     \epsftmp=\epsfxsize
     \loop \advance\epsfrsize\epsfrsize \divide\epsftmp 2
     \ifnum\epsftmp>0
        \ifnum\epsfrsize<\epsftsize\else
           \advance\epsfrsize-\epsftsize \advance\epsfysize\epsftmp \fi
     \repeat
     \epsfrsize=0pt
    \else
     \epsfrsize=\epsfysize
    \fi
   \fi
   \ifepsfverbose\message{#1: width=\the\epsfxsize, height=\the\epsfysize}\fi
   \epsftmp=10\epsfxsize \divide\epsftmp\pspoints
   \vbox to\epsfysize{\vfil\hbox to\epsfxsize{%
      \ifnum\epsfrsize=0\relax
        \includegraphics{#1}%
      \else
        \epsfrsize=10\epsfysize \divide\epsfrsize\pspoints
        \includegraphics{#1}%
      \fi
      \hfil}}%
\global\epsfxsize=0pt\global\epsfysize=0pt}%
{\catcode`\%=12 \global\let\epsfpercent=
\long\def\epsfaux#1#2:#3\\{\ifx#1\epsfpercent
   \def\testit{#2}\ifx\testit\epsfbblit
      \epsfgrab #3 . . . \\%
      \epsffileokfalse
      \global\epsfbbfoundtrue
   \fi\else\ifx#1\par\else\epsffileokfalse\fi\fi}%
\def\epsfempty{}%
\def\epsfgrab #1 #2 #3 #4 #5\\{%
\global\def\epsfllx{#1}\ifx\epsfllx\epsfempty
      \epsfgrab #2 #3 #4 #5 .\\\else
   \global\def\epsflly{#2}%
   \global\def\epsfurx{#3}\global\def\epsfury{#4}\fi}%
\def\epsfsize#1#2{\epsfxsize}
\let\epsffile=\epsfbox

\def\Phim#1{\mathrel{\mathop{\kern0pt \Phi}\limits^#1}}
\def\Psim#1{\mathrel{\mathop{\kern0pt \Psi^*}\limits^#1}}
\def\phib{\overline{\varphi}}
\def\C{{\bf C}}
\def\Z{{\bf Z}}
\def\H{{\cal H}}
\def\A{{\cal A}}

\def\vep{\varepsilon}
\def\z{\zeta}

\def\be{\begin{equation}}
\def\en{\end{equation}}

\def\bea{\begin{eqnarray}}
\def\ena{\end{eqnarray}}
\def\bean{\begin{eqnarray*}}
\def\enan{\end{eqnarray*}}
\def\lb#1{\label{#1}}
\def\rf#1{(\ref{#1})}

\def\lbvac#1{{}_B\langle#1|}
\def\rbvac#1{|#1\rangle_B}
\def\C{{\bf C}}

\def\e{\epsilon}
\def\s{\sigma}

\def\bra#1{\langle #1 |}	
\def\ket#1{| #1\rangle}		
\def\br#1{\langle #1 \rangle}	

\def\XXZ{{X\hskip-2pt X\hskip-2pt Z}}
\def\XYZ{{X\hskip-2pt Y\hskip-2pt Z}}
\def\itm#1{\par\noindent\quad#1}
\def\aip(#1){(#1;p^2,q^4)_\infty}
\def\bip(#1){(#1;p^2,q^8)_\infty}
\def\snh{\hbox{\rm snh}}
\def\cip(#1){(#1;p,q^4)_\infty}

\def\path{|p\rangle}

\def\id{{\rm id}\,}

\def\Phit{\widetilde{\Phi}}  
\def\Psit{\widetilde{\Psi}}  

\def\Phim#1{\mathrel{\mathop{\kern0pt \Phi}\limits^#1}}
\def\Psim#1{\mathrel{\mathop{\kern0pt \Psi}\limits^#1}}
\def\Phin#1{\mathrel{\mathop{\kern0pt \Phit}\limits^#1}}
\def\Psin#1{\mathrel{\mathop{\kern0pt \Psit}\limits^#1}}

\catcode`@=11
\def\citen#1{%
\if@filesw \immediate \write \@auxout {\string \citation {#1}}\fi
\@tempcntb\m@ne \let\@h@ld\relax \def\@citea{}%
\@for \@citeb:=#1\do {%
  \@ifundefined {b@\@citeb}%
    {\@h@ld\@citea\@tempcntb\m@ne{\bf ?}%
    \@warning {Citation `\@citeb ' on page \thepage \space undefined}}%
    {\@tempcnta\@tempcntb \advance\@tempcnta\@ne
    \setbox\z@\hbox\bgroup 
    \ifnum0<0\csname b@\@citeb \endcsname \relax
       \egroup \@tempcntb\number\csname b@\@citeb \endcsname \relax
       \else \egroup \@tempcntb\m@ne \fi
    \ifnum\@tempcnta=\@tempcntb 
       \ifx\@h@ld\relax 
          \edef \@h@ld{\@citea\csname b@\@citeb\endcsname}%
       \else 
          \edef\@h@ld{\hbox{--}\penalty\@highpenalty
            \csname b@\@citeb\endcsname}%
       \fi
    \else   
       \@h@ld\@citea\csname b@\@citeb \endcsname
       \let\@h@ld\relax
    \fi}%
 \def\@citea{,\penalty\@highpenalty}%
}\@h@ld}
\def\@citex[#1]#2{\@cite{\citen{#2}}{#1}}%
\def\@cite#1#2{\leavevmode\unskip
  \ifnum\lastpenalty=\z@\penalty\@highpenalty\fi
  \ [{\multiply\@highpenalty 3 #1
      \if@tempswa,\penalty\@highpenalty\ #2\fi 
    }]\spacefactor\@m}
\catcode`@=12
\newcommand{\bq}{\begin{equation}}
\newcommand{\eq}{\end{equation}}
\newcommand{\bqa}{\begin{eqnarray}}
\newcommand{\eqa}{\end{eqnarray}}
\newcommand{\ra}{\rightarrow}

\def\G{\Gamma}
\def\half{{1 \over 2}}
\def\s{\sigma}

\def\D{\Delta}
\def\a{\alpha}

\def\p{\phi}
\def\et{\xi}
\def\etp{\xi^{\prime}}

\def\ov{\over}

\def\ed{\end{document}}
\def\ws{\;\;}

\def\ra{\rightarrow}

\def\2pi{1\over 2\pi i}

\def\newline{\hfil\break}

\def\ra{\rightarrow}

\def\sq2{\sqrt{2}}
\def\sqk2{\sqrt{2(k+2}}
\def\sqk{\sqrt{k}}

\def\pt{\phi^{\ast}}

\def\bfig{\begin{figure}}
\def\bfigt{\begin{figure}[top]}
\def\efig{\end{figure}}
\def\be{\begin{equation}}
\def\ee{\end{equation}}
\def\br{\begin{array}}
\def\er{\end{array}}
\def\bea{\begin{eqnarray}}
\def\eea{\end{eqnarray}}
\def\ba{\begin{equation}\begin{array}}
\def\ea{\end{array}\end{equation}}
\def\bac{\begin{equation}\begin{array}{rll}}

\def\Z{{\Bbb Z}}
\def\C{{\Bbb C}}

\def\cH{{\cal{H}}}
\def\cHb{\bar{{\cal{H}}}}

\def\cM{{\cal{M}}}

\def\sl{\sum\limits}

\def\zp{\zeta^{\prime}}

\def\vep{\varepsilon}

\def\c{\chi}
\def\Ab{\bar{A}}
\def\z{\zeta}

\begin{document}
\begin{flushright}RIMS-1005, CRM-2246\\
hep-th/9502060 \\[10mm]
\end{flushright}
\begin{center}
{\LARGE  Difference Equations in Spin Chains with a Boundary}\\[8mm]
{\large Michio Jimbo,$\!^1$ Rinat Kedem,$\!^2$ \\Hitoshi Konno,$\!^3$
Tetsuji Miwa$^2$ \\ and Robert Weston$^4$\\[8mm]
December 1994}
\end{center}
\footnotetext[1]{Department of Mathematics, Faculty of Science,
Kyoto University, Kyoto 606, Japan.}
\footnotetext[2]{Research Institute for Mathematical Sciences,
Kyoto University, Kyoto 606, Japan.}
\footnotetext[3]{Yukawa Institute for Theoretical Physics,
Kyoto University, Kyoto 606, Japan.}
\footnotetext[4]{Centre de Recherches Math\'ematiques,
Universit\'e de Montr\'eal,
C.P. 6128, Succursale Centre-Ville,  Montr\'eal (Qu\'ebec) H3C 3J7, Canada.}
\begin{abstract}
Correlation functions and form factors in vertex models or spin chains
are known to satisfy certain difference equations called the quantum
Knizhnik-Zamolodchikov equations. We find similar difference equations
for the case of semi-infinite spin chain systems with integrable
boundary conditions. We derive these equations using the properties of
the vertex operators and the boundary vacuum state, or alternatively
through corner transfer matrix arguments for the
8-vertex model with a
boundary. The spontaneous boundary magnetization is found by solving
such difference equations. The boundary $S$-matrix is also proposed
and compared, in the sine-Gordon limit, with Ghoshal--Zamolodchikov's
result.  The axioms satisfied by the form factors in the boundary
theory are formulated.
\end{abstract}
\setcounter{section}{0}
\setcounter{equation}{0}
\section{Introduction}
Following the success of conformal field theory in two dimensions,
much attention has recently been focused on trying to understand
the algebraic structure of solvable lattice models and massive
integrable quantum field theories \cite{FR,JM,Smir,BeLe}.
Certain quantum deformations of affine Lie algebras play a
central role in the description of these systems.
One of the remarkable results of these studies is the
discovery of the difference analogues of the
Knizhnik--Zamolodchikov (KZ) equation~\cite{KnZa}.
These `quantum KZ equations' are
satisfied by both the correlation functions and the form factors
of integrable models~\cite{FR,JM,Smir,Smi92}.
They allow us to study
the off-shell properties of the models.

In this paper we establish similar results for
the $\XYZ$ Hamiltonian on a semi-infinite chain, with an
interaction at the boundary corresponding to a magnetic field in the
$z$-direction:
\be
H_B=-\frac{1}{2}\sum_{k=1}^{\infty}\Bigl((1+\Gamma)
\sigma^x_{k+1}\sigma^x_{k}+(1-\Gamma)
\sigma^y_{k+1}\sigma^y_{k}+\Delta\sigma^z_{k+1}\sigma^z_{k}\Bigr)
+h\sigma^z_1.
\label{hamiltonian}
\en



Integrable models with special
boundary conditions which preserve integrability,
have been studied both in lattice and continuum theories,
e.g., \cite{MW,Car}. In general frameworks,
the boundary interaction is specified
by a reflection matrix $K$ for lattice systems \cite{Skl87}, or
by a boundary $S$-matrix for quantum field theories \cite{GZ}.
Integrability is guaranteed by the fact that the
they satisfy the boundary Yang-Baxter equation.
The Hamiltonian \rf{hamiltonian} is obtained from a commuting transfer matrix
constructed by such a $K$-matrix \cite{IK,HOU}. This model is
particularly interesting because the sine-Gordon model \cite{GZ}
is its continuum limit.

In a recent paper~\cite{JKKKM}, it is shown that the space of states of
the $\XXZ$ spin chain with a boundary can be described in terms of the
vertex operators $\Phi_\vep(\z),~\Psi^*_\vep(\z)$ associated with the
``bulk'' (infinite chain) $\XXZ$ model \cite{JM}.
Explicit expressions for the two vacuum
states $|i\rangle_B,~ (i=0,1)$ of the boundary $\XXZ$ Hamiltonian
are obtained by using the bosonization formula for the vertex operators,
and then they are used to obtain the boundary excited energy and the boundary
$S$-matrix.

It is an interesting problem to extend
the analysis of the off-shell quantities
such as the correlation functions and the form factors
in the bulk theories to the boundary cases.
In this paper, we show that the boundary analogues \cite{Cher}
of the quantum KZ equation
are satisfied by these quantities in the $\XXZ$ and $\XYZ$ models.
For the $\XXZ$ model, our results follow from the results in \cite{JKKKM}.
Assuming similar properties of the vertex operators \cite{FIJKMY}
and the vacuum states, we can extend the results to the $\XYZ$ model.
Alternatively, for the correlation functions we can
use corner transfer matrix arguments \cite{Bax82,JMN} to derive
the same difference equation.

%

In Section 2, we will derive the $q$-difference equations satisfied by
the correlation functions of the $\XXZ$ and $\XYZ$ spin chains with a
boundary. In Section 3, we derive the same equations for the $\XYZ$
chain using a graphical corner transfer matrix argument similar to
that used in \cite{JMN}.  In Section 4 we discuss the boundary
magnetization (the vacuum expectation value of the boundary spin
operator $\sigma_1^z$) of the $\XYZ$ model. In the case where $h=0$,
one can solve the corresponding difference equation, and the result is
simply the square of the bulk magnetization of the eight-vertex model.
A similar phenomenon was observed in the case of the $\XXZ$ model with
a boundary \cite{JKKKM}.  We conjecture the form of the boundary
$S$-matrix in Section 5. Lacking bosonization formulas for the vertex
operators of the $\XYZ$ model, we cannot derive it explicitly.
However, our result reduces to the appropriate sine-Gordon boundary
$S$-matrix of \cite{GZ} in the scaling limit. Finally, in Section 6, we present
the
analog of Smirnov's axioms \cite{Smi92} for the form factors.

\setcounter{section}{1}
\setcounter{equation}{0}
\def\refeq#1{\rf#1}
\setcounter{equation}{0}
\section{The boundary $q$-difference equation}
In this section, we derive the $q$-difference equation satisfied by
the boundary $n$-point function, i.e. the expectation value of the
product of type I vertex operators between the vacuum states of the
boundary Hamiltonian.  This equation was first discovered by Cherednik
\cite{Cher}.  We will derive this equation first for the boundary
$\XXZ$ model in the formulation of \cite{JKKKM}, and then extend it to
the boundary $\XYZ$ model.

\subsection{The $\XXZ$ model}

The boundary $\XXZ$ model was formulated in \cite{JKKKM} in terms of
the vertex operators of $U_q(\widehat{sl}_2)$.
Let us recall the basic relations for type I vertex operators
$\Phi_{\vep}(\z)$ and $\Phi^*_{\vep}(\z)=\Phi_{-\vep}(-q^{-1}\z),~\vep=\pm$
\cite{JM},
and the boundary ground state $|0\rangle_B$ of the $\XXZ$ Hamiltonian
defined in
\cite{JKKKM}. For the definition of the $R$-matrix $R(\z)$ and the
boundary reflection matrix $K(\z)$ we refer the reader to \cite{JKKKM} (see
(A.1) and (A.2), and (2.2) and (2.3) therein, and see also (A.11) for the
scalar $g$
which appears below). The type I vertex operators commute as
\begin{equation}
\sum_{\vep'_1,\vep'_2}R^{\vep'_1,\vep'_2}_{\vep_1,\vep_2}(\z_1/\z_2)
\Phi_{\vep'_1}(\z_1)\Phi_{\vep'_2}(\z_2)=
\Phi_{\vep_2}(\z_2)\Phi_{\vep_1}(\z_1)~,\lb{CR}
\end{equation}
and have the inversion property
\begin{equation}
g~\Phi_{\vep_1}(\z)\Phi^*_{\vep_2}(\z)=\delta_{\vep_1\vep_2}\hbox{\rm id}~.
\lb{INV}
\end{equation}
The boundary ground state and its dual have the following reflection
properties with respect to type I vertex operators:
\begin{eqnarray}
&&\sum_{\vep'}K^{\vep'}_\vep(\z)\Phi_{\vep'}(\z)|0\rangle_B
=\Phi_{\vep}(\z^{-1})|0\rangle_B~,\lb{RBS}\\
&&\sum_{\vep'}\lbvac0\Phi^*_{\vep'}(\z^{-1})K^{\vep}_{\vep'}(\z)
=\lbvac0\Phi^*_{\vep}(\z)~.\lb{LBS}
\end{eqnarray}

The relation between these and the spin chain Hamiltonian will be discussed
in the context of the $\XYZ$ model in Section 2.3. Here we note only that
the (renormalized) $\XXZ$ Hamiltonian is given by
 $H=const.dT_B(\z)/d\z\Bigl|_{\z=1}$, where the transfer matrix is defined by
\be
T_B(\z)=g\sum_{\vep,\vep'}
\Phi^*_\vep(\z^{-1})K^{\vep'}_\vep(\z)\Phi_{\vep'}(\z).\lb{TB}
\en
The ground state $\ket{0}_B$ satisfies $T_B(\z)\ket{0}_B=\ket{0}_B$.

Consider the boundary $n$-point function,
\begin{equation}
G(\z_1,\cdots,\z_n)=
\sum_{\vep_1,\cdots,\vep_n}\lbvac0\Phi_{\vep_1}(\z_1)
\cdots\Phi_{\vep_n}(\z_n)\rbvac0
v^{\vep_1}\otimes\cdots\otimes v^{\vep_n},
\label{eqn:Gn}
\end{equation}
where $v^{\pm}$ are the weight vectors in the $2$-dimensional vector space
$V=\C v^+\oplus\C v^-$.
Note that $R(\z)\in\hbox{\rm End}_\C(V\otimes V)$ and
$K(\z)\in\hbox{\rm End}_\C(V)$.
Equations \rf{CR}, \rf{RBS} and \rf{LBS}
imply
\bean
&&R_{j\,j+1}(\z_j/\z_{j+1})~G(\z_1,\cdots,\z_j,\z_{j+1},\cdots,\z_n)
=P_{j\,j+1}G(\z_1,\cdots,\z_{j+1},\z_j,\cdots,\z_n),\\
&&K_n(\z_n)~G(\z_1,\cdots,\z_n)=G(\z_1,\cdots,\z_n^{-1}),\\
&&\overline{K}_1(\z_1)~G(\z^{-1}_1,\cdots,\z_n)=
G(q^{-2}\z_1,\z_2,\cdots,\z_n).
\enan
Here,
\bean
&&P_{j\,j+1}
(v^{\vep_1}\otimes\cdots\otimes v^{\vep_{j+1}}\otimes v^{\vep_j}\otimes\cdots
\otimes v^{\vep_n})
=v^{\vep_1}\otimes\cdots\otimes v^{\vep_j}\otimes v^{\vep_{j+1}}\otimes\cdots
\otimes v^{\vep_n},\\
&&R_{j\,j+1}(\z)(v^{\vep_1}\otimes\cdots\otimes v^{\vep_n})
\quad=\sum_{\vep'_j,\vep'_{j+1}}
R^{\vep_j,\vep_{j+1}}_{\vep'_j,\vep'_{j+1}}(\z)
(v^{\vep_1}\otimes\cdots\otimes v^{\vep'_j}
\otimes v^{\vep'_{j+1}}\otimes\cdots\otimes v^{\vep_n}),\\
&&K_j(\z)(v^{\vep_1}\otimes\cdots\otimes v^{\vep_n})
=\sum_{\vep'_j}K^{\vep_j}_{\vep'_j}(\z)
(v^{\vep_1}\otimes\cdots\otimes v^{\vep'_j}\otimes\cdots\otimes v^{\vep_n}),\\
&&\overline{K}_j(\z)(v^{\vep_1}\otimes\cdots\otimes v^{\vep_n})
=\sum_{\vep'_j}K^{-\vep'_j}_{-\vep_j}(-q^{-1}\z)
(v^{\vep_1}\otimes\cdots\otimes v^{\vep'_j}\otimes\cdots\otimes v^{\vep_n}).
\enan
{}From these equations we obtain
\bea
G(\z_1,\cdots,q^{-2}\z_j,\cdots,\z_n)
&=&R_{j\,j-1}(\z_j/q^2\z_{j-1})\cdots R_{j1}(\z_j/q^2\z_1)
\overline{K}_j(\z_j)\times\nonumber\\
&\times& R_{1j}(\z_1\z_j)\cdots R_{j-1\,j}(\z_{j-1}\z_j)
R_{j+1\,j}(\z_{j+1}\z_j)\cdots R_{nj}(\z_n\z_j)\times\nonumber\\
&\times& K_j(\z_j)R_{jn}(\z_j/\z_n)\cdots R_{j\,j+1}(\z_j/\z_{j+1})
G(\z_1,\cdots,\z_n).\lb{CHERDIFF}
\ena
This is a version of Cherednik's equation \cite{Cher}.

\subsection{Consistency conditions}

We have shown that the boundary $n$-point function
$G(\z_1,\cdots,\z_n)$ satisfies \rf{CHERDIFF} if
relations \rf{CR}, \rf{RBS} and  \rf{LBS} are satisfied,
without using any further details of the $\XXZ$ model. Therefore,
if we have more general settings for these relations,
we get more general solutions to equation \rf{CHERDIFF}.
We remark that these three relations along with \rf{INV}
imply the following consistency conditions for $R(\z)$ and $K(\z)$:
\itm{(i)}
Yang--Baxter equation
\be
R_{12}(\z_1/\z_2)R_{13}(\z_1/\z_3)R_{23}(\z_2/\z_3)
=R_{23}(\z_2/\z_3)R_{13}(\z_1/\z_3)R_{12}(\z_1/\z_2).\lb{YBE}
\en
\itm{(ii)}
Unitarity relation
\be
R_{12}(\z_1/\z_2)R_{21}(\z_2/\z_1)=\hbox{\rm id}.\lb{Un}
\en
\itm{(iii)}
Crossing relation
\be
R^{\vep_1\vep_2}_{\vep'_1\vep'_2}(-q^{-1}\z^{-1})
=R^{\vep_2\,-\vep'_1}_{\vep'_2\,-\vep_1}(\z).\lb{Cr}
\en
\itm{(iv)}
Boundary Yang--Baxter equation
\be
K_2(\z_2)R_{21}(\z_1\z_2)K_1(\z_1)R_{12}(\z_1/\z_2)
=R_{21}(\z_1/\z_2)K_1(\z_1)R_{12}(\z_1\z_2)K_2(\z_2).\lb{BYBE}
\en
\itm{(v)}
Boundary unitarity relation
\be
K(\z)K(\z^{-1})=\hbox{\rm id}.\lb{BU}
\en
\itm{(vi)}
Boundary crossing relation
\be
K^b_a(\z)=\sum_{a',b'}R^{b,\,-a'}_{-b'\,a}(\z^2)K^{a'}_{b'}(-q^{-1}\z^{-1}).
\lb{BX}
\en

For completeness, we add a few remarks following the results in \cite{JKKKM}.
The properties of the vertex operators
entails the following relations for (\ref{TB}): $[T_B(\z_1),T_B(\z_2)]=0$,
$T_B(1)=1$, $T_B(\z)T_B(\z^{-1})=1$, $T_B(-q^{-1}\z^{-1})=T_B(\z)$.
Therefore, if $t(\z)$ is an eigenvalue of $T_B(\z)$,
we have $t(1)=1$, $t(\z)t(\z^{-1})=1$, and
$t(-q^{-1}\z^{-1})=t(\z)$. A vector $|v\rangle$ (or $\langle v|$)
satisfies $\sum_{\vep'}K^{\vep'}_\vep(\z)\Phi_{\vep'}(\z)|v\rangle
=t(\z)\Phi_{\vep}(\z^{-1})|v\rangle$ (or $\sum_{\vep}\langle
v|\Phi^*_{\vep}(\z^{-1})K^{\vep'}_\vep(\z) =t(\z)\langle
v|\Phi^*_{\vep'}(\z)$) if and only if it is an eigenvector of
$T_B(\z)$ with the eigenvalue $t(\z)$. In this situation, the
$n$-point function
\be
G(\z_1,\cdots,\z_n)=
\sum_{\vep_1,\cdots,\vep_n}
\langle v|\Phi_{\vep_1}(\z_1)\cdots\Phi_{\vep_n}(\z_n)|v\rangle
v^{\vep_1}\otimes\cdots\otimes v^{\vep_n}\lb{n-point}
\en
satisfies the same equation \rf{CHERDIFF} because the effects of
$t(\z)$ in $\overline{K}(\z)$ and $K(\z)$ cancel. In particular,
$\lbvac1\Phi_{\vep_1}(\z_1)\cdots\Phi_{\vep_n}(\z_n)\rbvac1$ satisfies
\rf{CHERDIFF} (see 3.1 in \cite{JKKKM}).

\subsection{The $\XYZ$ model}

In this section we consider \rf{CHERDIFF} in the context of
the $\XYZ$ model, extending the setting of Section 2.1.  Unlike the
$\XXZ$ case, we do not have a complete solution for the $\XYZ$ model,
because we do not have the corresponding mathematical machinery, in
particular, a bosonization scheme. Certain results can, however, be
generalized. Following \cite{FIJKMY} we assume the existence of vertex
operators. We further assume the existence of the boundary vacuum
states, and derive \rf{CHERDIFF} for the boundary $n$-point
function. See, however, Section 3, where we give a physical argument
that supports the existence of such a setting.

To define the model, we use the variables
\be
p=e^{-{\pi K'\over K}},
\qquad -q=e^{-{\pi\lambda\over2K}},
\qquad\z=e^{{\pi u\over2K}},\qquad
z=\z^2,
\label{addp}
\en
where $K$, $K'$ are $I$, $I'$ in Baxter's notation~\cite{Bax82}, and
$\lambda$ and $u$ are as in (10.4.21) of~\cite{Bax82}. The variable $p$
is the elliptic nome, and $p=0$ corresponds to the $\XXZ$ case.
 We restrict
our discussion to the principal regime, $0<p^{1/2}<-q<\z^{-1}<1$.
Let
\begin{eqnarray*}
&&\hbox{\rm snh}(u)=-i\hbox{\rm sn}(iu),\quad
\hbox{\rm cnh}(u)=\hbox{\rm cn}(iu),\quad
\hbox{\rm dnh}(u)=\hbox{\rm dn}(iu),\\
&&(z;q_1,\cdots,q_m)_\infty
=\prod_{n_1,\ldots,n_m=0}^\infty(1-q_1^{n_1}\cdots q_m^{n_m}z),\quad
\Theta_q(z)=(z;q)_\infty(qz^{-1};q)_\infty(q;q)_\infty.
\end{eqnarray*}
The elliptic functions $\hbox{\rm sn}(u)$ etc., are found in the
appendix of \cite{Bax82}.  Unless otherwise stated, the elliptic nome
for these functions is $p$, and the corresponding modulus and conjugate modulus
are denoted by $k$ and $k'$. In \rf{energy} etc., we use elliptic
functions with elliptic nome
\be
-q=e^{-{\pi I'\over I}},
\label{newnome}
\en
modulus $k_I$ and conjugate modulus $k'_I$.

We first give solutions $R$ and $K$ to the consistency conditions
(i)--(vi).  The solution $R(\z)$ to equations (i)--(iii) for the
$\XYZ$ model is
\be
R(\z)=\pmatrix{a(\z)&&&d(\z)\cr
&b(\z)&c(\z)&\cr
&c(\z)&b(\z)&\cr
d(\z)&&&a(\z)\cr}~,\lb{xyzR}
\en
where the Boltzmann weights are
\[
a(\z)={1\over\mu(\z)}{\snh(\lambda-u)\over\snh(\lambda)},\quad
b(\z)={1\over\mu(\z)}{\snh(u)\over\snh(\lambda)},\quad
c(\z)={1\over\mu(\z)},\quad
d(\z)={k\over\mu(\z)}{\snh(\lambda-u)\snh(u)\over\snh(\lambda)},
\]
with
\[
{1\over\mu(\z)}=\frac{1}{\overline{\kappa}(\z^2)}
{(p^2;p^2)^2_\infty\over(p;p)^2_\infty}
{\Theta_{p^2}(q^2)\Theta_{p^2}(pz)\over\Theta_{p^2}(q^2z)},
\]
\be
\frac{1}{\overline{\kappa}(z)}
={\cip(q^4z^{-1})\cip(q^2z)\cip(pz^{-1})\cip(pq^2z)
\over\cip(q^4z)\cip(q^2z^{-1})\cip(pz)\cip(pq^2z^{-1})}.\lb{kappabar}
\en
Usually, will not denote the $p$ and $q$ dependence explicitly in cases when
no confusion should arise as a result of their absence.

The general $3$-parameter family of solutions $K(\z)$ to \rf{BYBE} was given
in \cite{IK} and \cite{HOU}. In the following, we restrict to the
diagonal solution $K(\z)$ with a single parameter
\be
r={\rm e}^{\pi\eta\over K}.
\label{addr}
\en

Let
\be
K(\z)=
{1\over f(\z;r)}{\widehat K}(\z;r),\qquad
{\widehat K}(\z;r)=
\pmatrix{{\snh(\eta+u)\over\snh(\eta-u)}&\cr&1\cr},\lb{KMAT}
\en
where
\begin{eqnarray*}
{1\over f(\z;r)}&=&
{\varphi(z;r)\over\varphi(z^{-1};r)},\\
\varphi(z;r)&=&{\aip(prz)\aip(p^2r^{-1}z)\aip(rq^4z)\aip(pr^{-1}q^4z)
\over\aip(prq^2z)\aip(p^2r^{-1}q^2z)\aip(rq^2z)\aip(pr^{-1}q^2z)}\\
&&\times{\bip(q^6z^2)\bip(pq^2z^2)\bip(pq^6z^2)\bip(p^2q^2z^2)\over
\bip(q^8z^2)\bip(pq^4z^2)^2\bip(p^2z^2)}.
\end{eqnarray*}
The scalar factor $f(\z;r)$ is chosen so that
\rf{BU}, \rf{BX} hold, and $K^-_-(\z)$
is analytic in the region $-q<|r|^{1/2}<\z^{-1}$.
In terms of the additive parameters defined in (\ref{addp}) and
(\ref{addr}), the region we consider is given by $0< u< -\eta<
\lambda< K'$.

Now let us recall the vertex operators defined for the elliptic
algebra corresponding to the $R$-matrix \rf{xyzR} in
\cite{FIJKMY}\footnote{In \cite{JMN}, $\Phi_\vep(\z)$ was introduced satisfying
\rf{CR}, and \rf{INV} with a different normalization, i.e.,
$g=1$. In this paper we follow the normalization given in \cite{FIJKMY}.}.
The type I vertex operator
\[
\Phi_\vep(\z):\H^{(i)}\rightarrow\H^{(1-i)}
\]
acts on graded vector spaces
\[
\H^{(i)}=\oplus_{d=0}^\infty\H^{(i)}_d\quad(i=0,1)
\]
with character
\[
\sum_{d=0}^\infty\hbox{\rm dim}\,\H^{(i)}_dt^d
={1\over\prod_{n=0}^\infty(1-t^{2n+1})}.
\]
The vertex operator is expanded in the form
\[
\Phi_\vep(\z)=\sum_{n\in\Z}\Phi_{\vep,n}\z^{-n}
=(-1)^{i+(1+\vep)/2}\Phi_\vep(-\z),
\]
where $\Phi_{\vep,n}:\H^{(i)}_d\rightarrow\H^{(i+1)}_{d-n}$.
The normalizations are such that $\Phi_{-,0}|0\rangle=|1\rangle$,
and $\Phi_{+,0}|1\rangle=|0\rangle$, where $\H^{(i)}_0=\C|i\rangle$.
In this normalization, we find that Eq. \rf{CR} holds with the
elliptic $R$-matrix \rf{xyzR}, and
Eq. \rf{INV} holds with \cite{FIJKMY}
\be
g={(pq^2;p)_\infty(pq^6;p,q^4)_\infty(pq^2;p,q^4)_\infty
(q^2;q^4)_\infty\over
(p;p)_\infty(pq^4;p,q^4)_\infty^2(q^4;q^4)_\infty}.\lb{gxyz}
\en

We now turn to the boundary $\XYZ$ hamiltonian. In analogy with the
$\XXZ$ case, we now assume that there exists a vacuum vector $\rbvac0$
(resp. $\lbvac0$) in an appropriate completion of $\oplus_{d:\hbox{\rm
\small even}}\H^{(0)}_d$, (resp. $\oplus_{d:\hbox{\rm \small
even}}\H^{(0)*}_d$), which satisfies \rf{RBS} (resp. \rf{LBS}).  In
Section 3, we give a graphical argument for the construction of these
states.  However, since we do not have a bosonization formula for the vertex
operators, we cannot give an explicit bosonic construction of
this state, as we did in \cite{JKKKM} for the $\XXZ$ Hamiltonian.

In the $\XXZ$ model, we found that there exists another vacuum
state, the boundary bound state $\rbvac1$, with an excitation energy
\be
\Lambda(\z;r)=
{K^\vep_\vep(\z;r)\over K^{-\vep}_{-\vep}(\z;r^{-1})}
={1\over\z^2}{\Theta_{q^4}(r\z^2)\Theta_{q^4}(q^2r\z^{-2})
\over\Theta_{q^4}(r\z^{-2})\Theta_{q^4}(q^2r\z^2)}.
\lb{ev1}
\en
It turns out that this formula remains valid without any change for
the elliptic $K(\z)$ given by \rf{KMAT}.  Therefore, for $|r|\ge |q|$,
we assume that there exists a vector $\rbvac1$ (resp. $\lbvac1$) in an
appropriate completion of $\oplus_{d:\hbox{\rm \small
even}}\H^{(1)}_d$, (resp. $\oplus_{d:\hbox{\rm \small
even}}\H^{(1)*}_d$), which satisfies
\begin{eqnarray}
&&\sum_{\vep'}K^{\vep'}_\vep(\z)\Phi_{\vep'}(\z)|1\rangle_B
=\Lambda(\z;r)\Phi_{\vep}(\z^{-1})|1\rangle_B~,\lb{RBS2}\\
&&\sum_{\vep'}\lbvac1\Phi^*_{\vep'}(\z^{-1})K^{\vep}_{\vep'}(\z)
=\Lambda(\z;r)\lbvac1\Phi^*_{\vep}(\z)~.\lb{LBS2}
\end{eqnarray}

Under these assumptions, we can formulate the boundary $\XYZ$ model
in an analogous way to the boundary $\XXZ$ model. In particular, the
correlation function
\[
\sum_{\vep_1,\cdots,\vep_n}\lbvac i\Phi_{\vep_1}(\z_1)
\cdots\Phi_{\vep_n}(\z_n)\rbvac i
v^{\vep_1}\otimes\cdots\otimes v^{\vep_n}
\]
with elliptic vertex operators satisfies the $q$-difference equation
\rf{CHERDIFF}.

\medskip

We define the transfer matrix of the boundary theory as in \rf{TB}
with the appropriate elliptic generalizations of $g,~K$ and
$\Phi_\vep(\z)$. The renormalized form of the Hamiltonian
\rf{hamiltonian} is defined from this transfer matrix by
\be
H_B^{\rm renor}=-\frac{\pi {\rm snh}(\lambda,k)}{4K}
\z\frac{d}{d\z}T_B(\z)\Bigl|_{\z=1}.
\label{renham}
\en
We find that the parameters in equation \rf{hamiltonian} are related
to the elliptic parameters above by
\be
\Gamma=k\ {\rm snh}^2(\lambda,k),\qquad
\Delta=-{\rm cnh}(\lambda,k){\rm dnh}(\lambda,k),\qquad
h=-{\rm snh}(\lambda,k)\frac{{\rm cnh}(\eta,k){\rm dnh}(\eta,k)}
{2{\rm snh}(\eta,k)}.
\label{hfield}
\en
This Hamiltonian is normalized so that its lowest eigenvalue is $0$.
{}From (\ref{ev1}) and (\ref{renham}),
the eigenvalues $e^{(i)}(r)\quad(i=0,1)$ of $H_B$ on $|i\rangle_B$ are
\bea
e^{(0)}(r)&=&0,\\
e^{(1)}(r)&=&\cases{\epsilon(1){\rm sn}(-2I'(\eta+iK)/\lambda,k_I')
&if $-1\leq r<-|q|$;\cr
{\displaystyle{\epsilon(1)\over k_I'{\rm sn}
(-2I'\eta/\lambda,k_I')}}
&if $ |q|<r<1$.\cr}\label{energy}
\ena
where we have
defined the following function
\bean
\epsilon(\xi)&=&
\frac{I}{K} {\rm snh}(\lambda,k) {\rm dn}(\frac{2I}{\pi}\theta,k_I),
\qquad \xi=-ie^{i\theta},\\
\epsilon(1)&=&
\frac{Ik_I'}{K} {\rm snh}(\lambda,k).
\enan
The reader should be careful about the nome of the elliptic functions.
(See (\ref{newnome}).)
The function $\epsilon(\xi)$ differs from the one in the
$\XXZ$ chain\cite{JKKKM} due to the difference of the normalization
factor in front of the derivative of $T_B(\z)$ in (\ref{renham}).
It reduces to the $\XXZ$ excitation energy when $p=0$.

The excited states are created by the action of type II vertex operators
$\Psi^*_{\mu}(\xi) (\mu=\pm)$ \cite{FIJKMY} on the vacuum vectors $\ket{i}_B$:
\be
\Psi^*_{\mu_m}(\xi_m)\cdots\Psi^*_{\mu_1}(\xi_1)\ket{i}_B.
\en
The energy spectrum of the one particle excitation is evaluated as
$e^{(i)}(r)+\epsilon(\xi)$. The excitation is therefore massive with
 mass $\epsilon(1)$ \cite{JKM}.
\setcounter{equation}{0}
\setcounter{section}{2}
\section{The Corner Transfer Matrix}
In this section,
we show how within the context of an inhomogeneous boundary vertex
model  it is possible to construct lattice realizations
of $\Phi_{\vep}(\z)$, $\Phi_{\vep}^{\ast}(\z)$,
$\ket{0}_B$ and ${_B{\bra{0}}}$ that obey relations
\rf{CR}, \rf{RBS} and \rf{LBS}.
We then present a rather physical picture of the origin of
the difference equations for correlation functions of local
operators in this vertex model.
\subsection{The Partition Function}
Following Sklyanin \cite{Skl87}, we build a lattice
with $2M$ vertical and $2N$ horizontal lines
as in Figure 1.
\bfigt\vspace*{-10mm}\hspace*{15mm}\epsfysize=150mm \epsffile{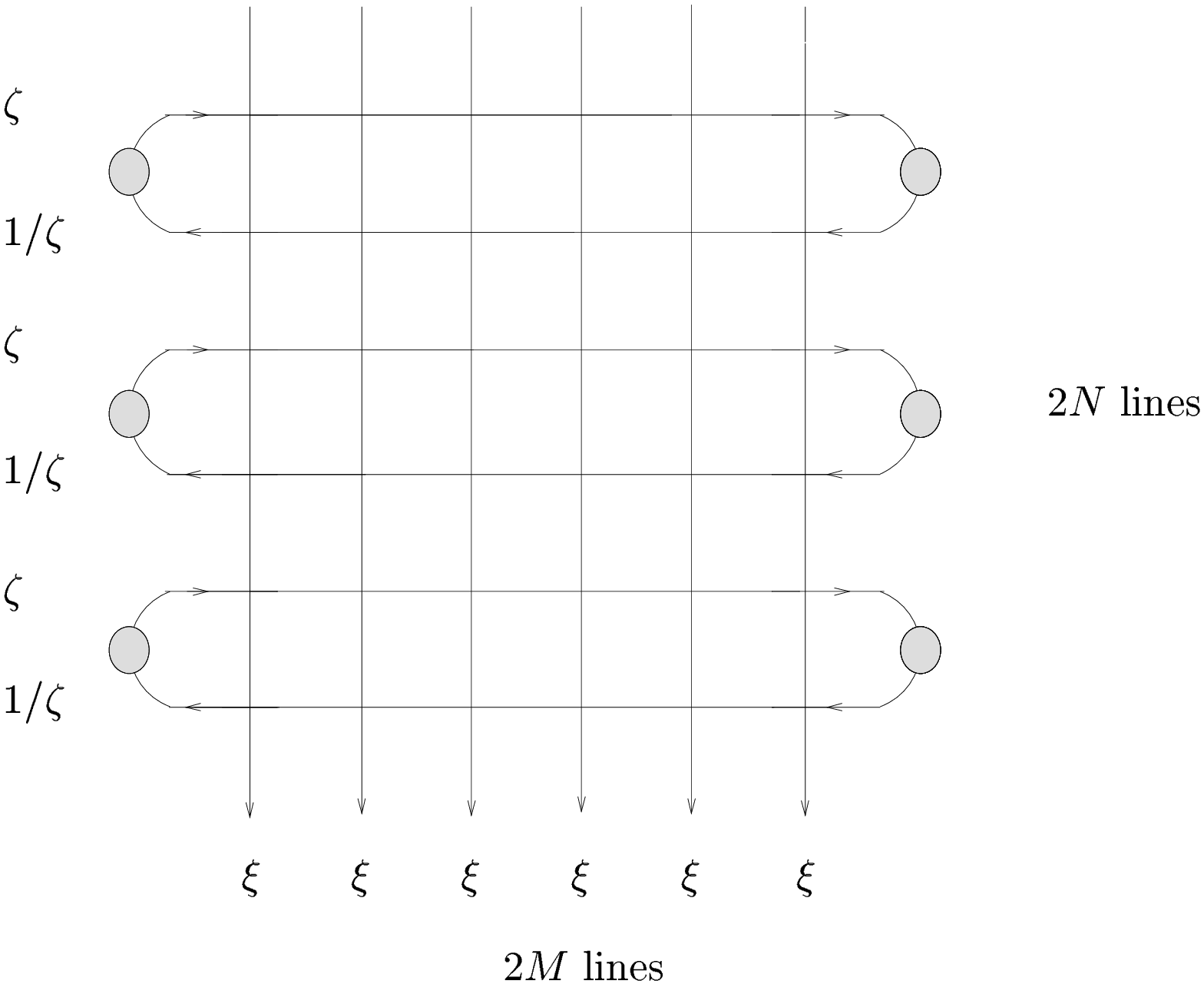}%
\vspace*{-50mm}
\caption{The inhomogeneous $2M \times 2 N$ lattice ${\cal L}_{MN}$.
The circles on the right hand side
denote reflection matrices $K(\z)$, and those on the left hand side
denote $K(-q^{-1}\z^{-1})^t$.}\efig
Here, in the vertical direction we impose the cyclic boundary condition.
Let us call this lattice ${\cal L}_{MN}$.
Let the horizontal lines carry alternating spectral parameters $\z$ and
$1/\z$, and the vertical lines $\et$. We consider the region
$0<p^{1/2}<-q<|r|^{1/2}<|\z/\xi|^{-1},|\z\xi|^{-1}<1$.
We use the word ground state to refer to the lowest energy
states in the $h=0$, $\Gamma=0$, $\D \ra -\infty$ limit.
There are two such ground states, one of which is shown in Figure 2.
\bfigt\vspace*{-15mm}\hspace*{15mm}\epsfysize=165mm
\epsfbox{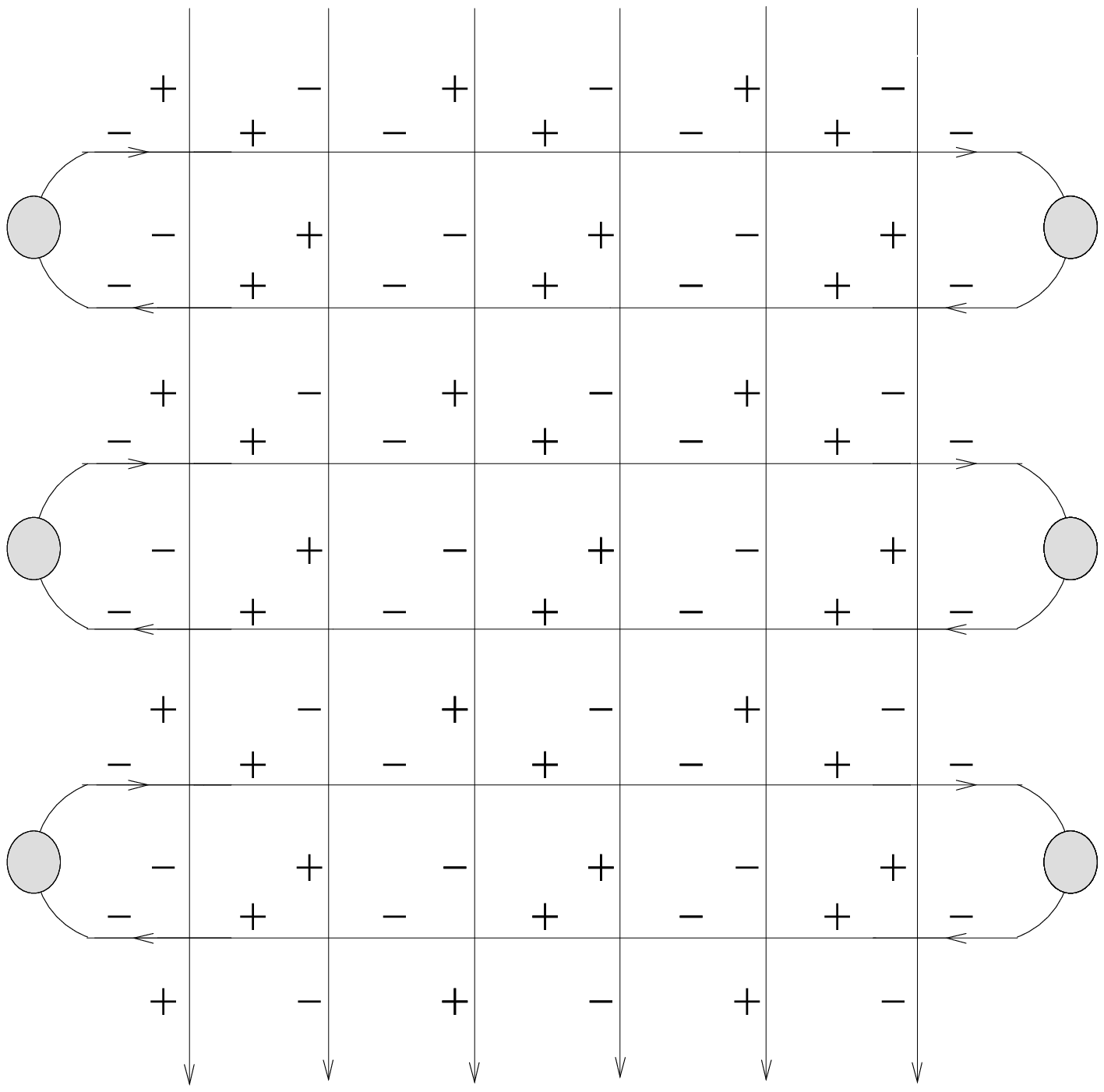}\vspace*{-70mm}%
\caption{The $i=0$ ground state}\efig
Note that if $h\ne0$ then the degeneracy of the ground state energies
is resolved, and only one of the ground states corresponds to the
lowest energy. Nevertheless, we call both of them ground states
for all values of $h$. Corresponding to these two ground states,
we have two antiferromagnetic sectors.
Let us label these sectors by $i=0,1$.
When we consider the partition function and correlation functions,
we choose one of the ground state sectors,
and take configuration sums over such states that are
different from the ground state at finitely many edges.

Let us restrict to the region $h\ge0$. We conjecture that
in the limit $M,N\rightarrow\infty$
the partition function $Z^{(i)}_{MN}$ of this lattice behaves as
\[
\log Z^{(i)}_{MN} ~~\sim ~~
MN\Bigl(\log \mu^{(i)}(\z/\et)+\log \mu^{(i)}(\z\et)\Bigr) +
N\Bigl(\log\nu^{(i)}(\z)+\log\nu^{(i)}(-q^{-1}\z^{-1})\Bigr).
\]
Here $\mu^{(i)}$ is the partition function per site in the bulk theory,
and $\nu^{(i)}$ is the partition function per boundary site,
which in the present normalization are given by
\begin{eqnarray}
&&\mu^{(i)}(\z)=1 \qquad \hbox{for $i=0,1$},
\\
&&\nu^{(0)}(\z)=1, \qquad \nu^{(1)}(\z)=\Lambda(\z;r).
\label{eqn:nu1}
\end{eqnarray}
These conjectures have been suggested
by an argument similar to the inversion trick in the bulk theory,
on the basis of the properties \rf{YBE}--\rf{BX} and \rf{ev1} for
$R$ and $K$ matrices.
Here we will not discuss the details.

Now consider the lattice shown in Figure 3. We divide this lattice
into the  four sections
indicated by the dotted lines.
\bfigt\vspace*{-15mm}\hspace*{15mm}\epsfysize=220mm \epsffile{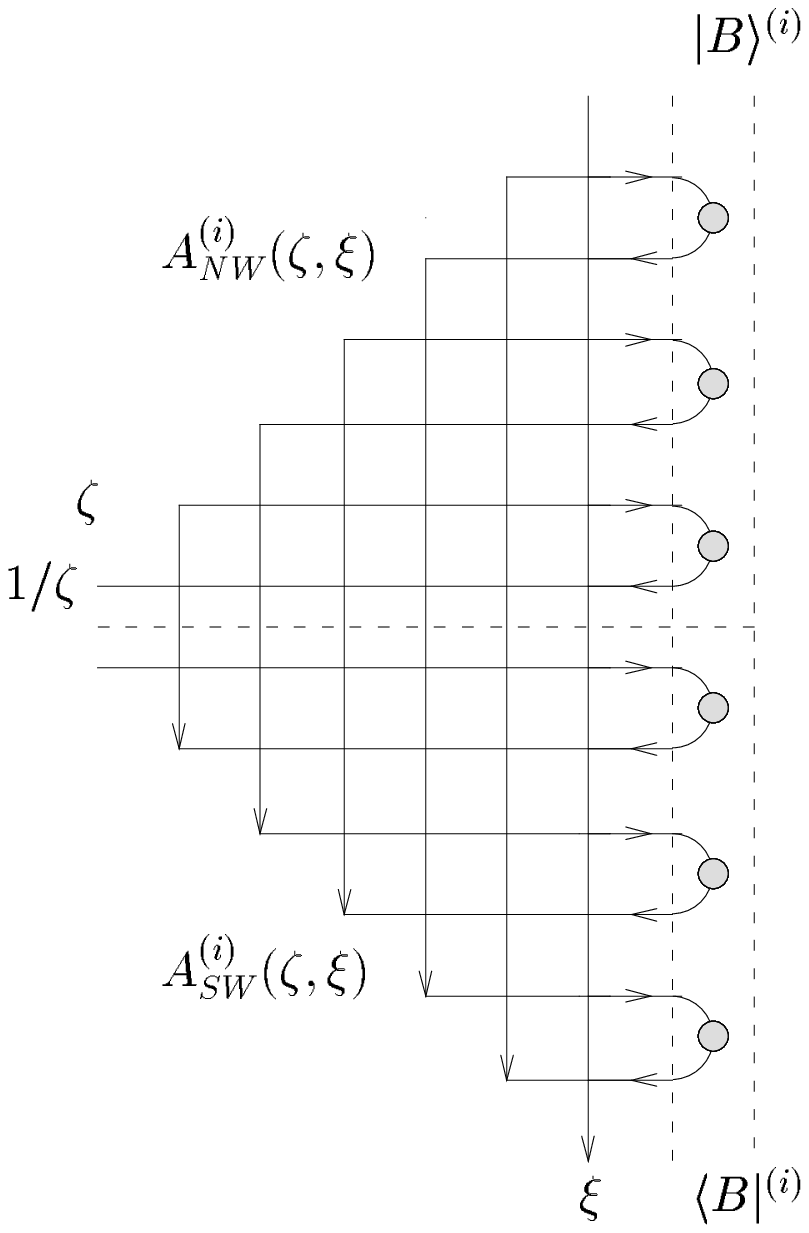}%
\vspace*{-105mm}
\caption{The inhomogeneous lattice split into four sections
 }\efig
We denote the NW and SW corner transfer matrices
(CTM's), by $A_{NW}^{(i)}(\z,\et)$ and $A_{SW}^{(i)}(\z,\et)$, and
the upper and lower lines of $K(\z)$ matrices by $\ket{B}^{(i)}$ and
${^{(i)}{\bra{B}}}$.
In the infinite lattice limit we can identify
$A_{NW}^{(i)}(\z,\et)$, $A_{SW}^{(i)}(\z,\et)$, $\ket{B}^{(i)}$ and
${}^{(i)}\bra{B}$ as maps between, or elements of,
the vectors spaces $\cHb^{(i)}$ and $\cH^{(i)}$ (and their duals
$\cHb^{\ast\,(i)}$ and $\cH^{\ast\,(i)}$).
These vector spaces are defined as the spans of the half-infinite
pure tensor vectors
\bean
\cdots \otimes v_{p(3)}\otimes v_{p(2)} \otimes v_{p(1)},~~~
\rm{with}~~~ &&
 p(j) = (-)^{j+i},~~ j\gg 1,~~ \hbox{for}~~ \cH^{(i)}, \\
&& p(j) = (-)^{1+i},~~ j\gg 1,~~ \hbox{for} ~~ \cHb^{(i)}.
\enan
The identification is that
\begin{eqnarray*}
A_{NW}^{(i)}(\z,\et)&:& \cHb^{(i)} \ra \cH^{(i)},\\
A_{SW}^{(i)}(\z,\et)&:& \cH^{(i)} \ra \cHb^{(i)},\\
\ket{B}^{(i)}&\in & \cHb^{(i)},\\
{^{(i)}{\bra{B}}}&\in &\cHb^{\ast\,(i)}.
\end{eqnarray*}
The partition function of the lattice is then given by
\[
Z^{(i)}(\z,\et)
={^{(i)}{\bra{B}}}A_{SW}^{(i)}(\z,\et)A_{NW}^{(i)}(\z,\et)\ket{B}^{(i)}.
\]
\subsection{Vertex Operators}
We define `vertex operators' $\p_\vep^{(1-i,i)}(\et)$,
$\p_\vep^{*(1-i,i)} (\et)$,
$\p_{\vep}^{U,(1-i,i)}(\z,\et)$
and $\p_{\vep}^{D,(1-i,i)}(\z,\et)$ by the half-infinite
lattice insertions shown in Figure 4.
The superscripts $(1-i,i)$  indicate how the vertex operators act on
$\cH^{(i)}$ and $\cHb^{(i)}$. Namely,
\ba{rclrcl} \p_\vep^{(1-i,i)}(\et)&:& \cH^{(i)} \ra \cH^{(1-i)}
\ws&\ws \p_\vep^{*(1-i,i)} (\et) &:& \cH^{(i)} \ra \cH^{(1-i)}\\
\p_{\vep}^{U,(1-i,i)}(\z,\et)&:&\cHb^{(i)} \ra \cHb^{(1-i)} \ws&\ws
\p_{\vep}^{D,(1-i,i)}(\z,\et)&:&\cHb^{(i)} \ra \cHb^{(1-i)}
\label{vo1}.\ea
Henceforth these superscripts will be suppressed.
\bfigt\vspace*{-15mm}\epsfysize=180mm \epsfbox{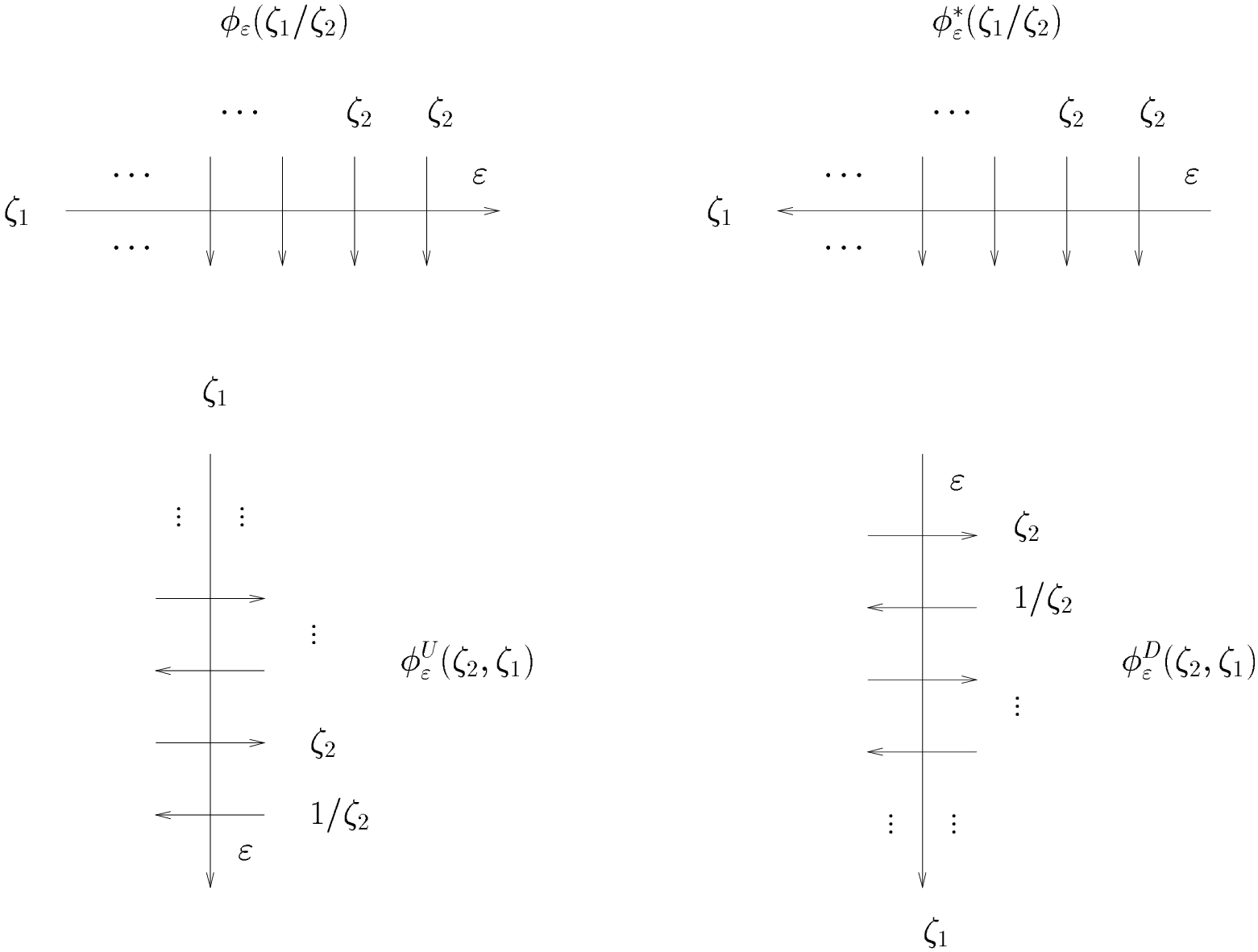}\vspace*{-60mm}%
\caption{The graphical definition of vertex operators}\efig\nopagebreak

We  now argue that the vertex operators \rf{vo1}
and CTM's obey
a set of relations that will allow us to use
them to construct lattice realizations of
$\Phi_{\vep}(\z)$, $\Phi_{\vep}^{\ast}(\z)$,
$\ket{0}_B$ and ${_B{\bra{0}}}$ obeying relations
\rf{CR}, \rf{RBS} and \rf{LBS}. These relations are
\bqa
\p_{\vep}(\et) &=&\pt_{-\vep}(-q \et), \label{cross}\\
\sl_{\vep'_1,\vep'_2} R_{\vep_1,\vep_2}^{\vep'_1,\vep'_2}
(\et_1/\et_2) \p_{\vep'_1}(\et_1)\p_{\vep'_2}(\et_2)
&=&\p_{\vep_2}(\et_2) \p_{\vep_1}(\et_1)
,\label{yb}\\
\sl_{\vep'} K_{\vep}^{\vep'}(\etp)
\p_{\vep'}^U(\z,\etp)\ket{B}^{(i)}
&=&\nu^{(i)}(\etp)
\p_{\vep}^U(\z,1/\etp) \ket{B}^{(i)} \label{btop}\\
\sl_{\vep'} {^{(i)}{\bra{B}}} \p_{\vep'}^D(\z,1/\etp)K_{\vep'}^{\vep}(\etp)
&=&\nu^{(i)}(\etp)
{^{(i)}{\bra{B}}} \p_{\vep}^D(\z,\etp)
,\label{bbot}\\
A_{NW}^{(i)}(\z,\et)\p_{\vep}^U(\z,\etp) &=& \p_{\vep}(\etp/\et)
A_{NW}^{(1-i)}(\z,\et)
,\label{ctmtop}\\
 A_{SW}^{(i)}(\z,\et) \pt_{\vep}(\etp/\et) &=&
\p_{\vep}^D(\z,\etp) A_{SW}^{(1-i)}(\z,\et) ,
\label{ctmbot}\eqa
where $\nu^{(i)}(\etp)$ are given in (\ref{eqn:nu1}).
Properties \rf{cross} and \rf{yb} are consequences
of the crossing symmetry of the $R$-matrix and the Yang-Baxter equation
(in the infinite lattice limit) respectively. They are discussed in reference
\cite{JMN}. The point in the argument for \rf{yb} is as follows.
If we compare the graphical expressions for
both sides of \rf{yb}, the left hand side contains one more site (or vertex),
i.e.,
$R_{\vep_1,\vep_2}^{\vep'_1,\vep'_2}(\et_1/\et_2)$, in the formula,
than the right hand site. The effect of the additional site is to multiply
by the partition function per site $\mu^{(i)}$. Since we normalized $R$ so that
$\mu^{(i)}=1$, we have \rf{yb}.
Equation \rf{btop} follows from a similar graphical argument.
See Figure 5.
\bfigt\vspace*{-15mm}\epsfysize=180mm \epsfbox{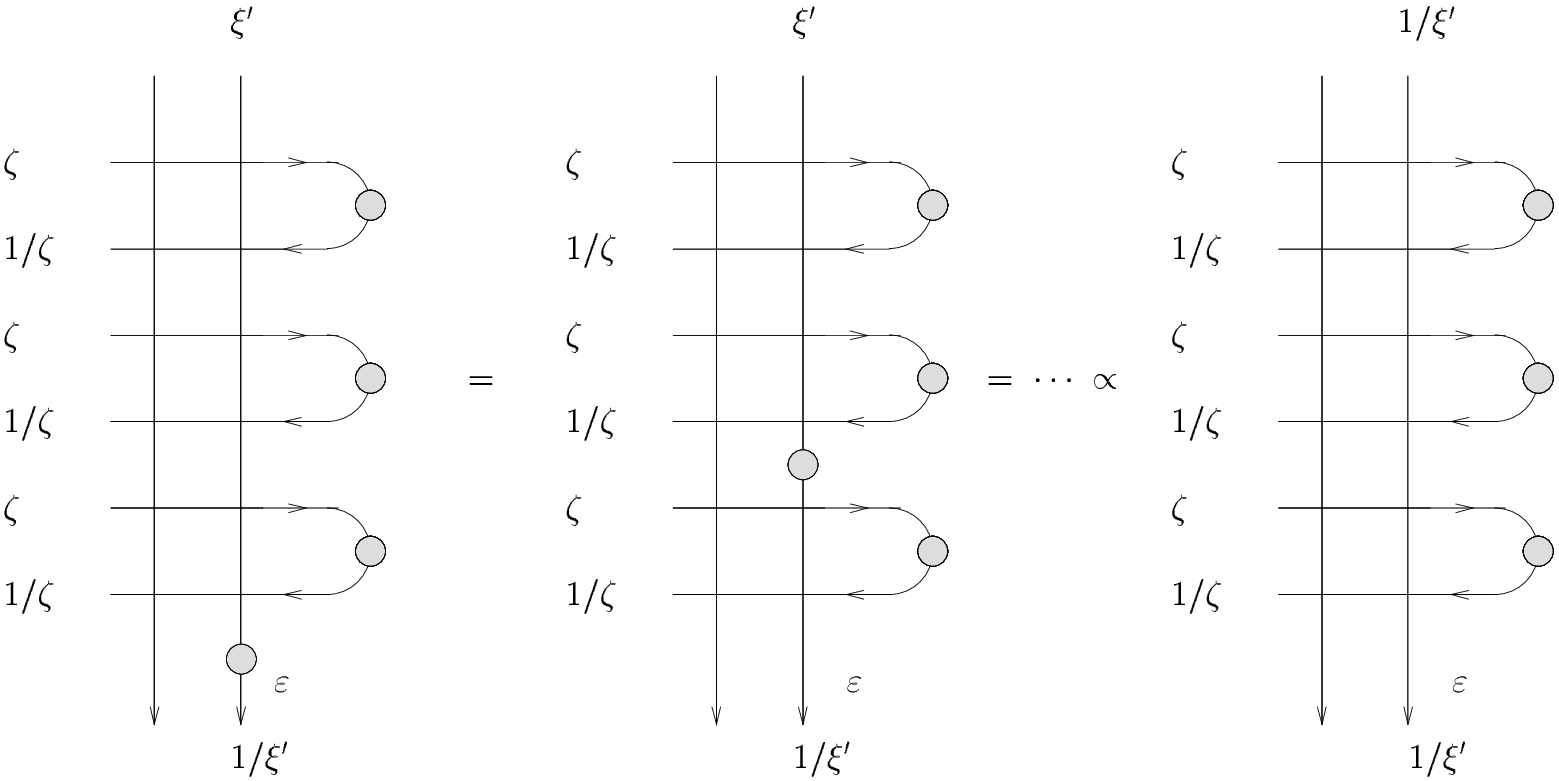}\vspace*{-95mm}%
\caption{The boundary reflection property}\efig
We use the boundary
Yang Baxter equation in order to move the $K(\etp)$
matrix up the boundary to infinity. This time we get the factor $\nu^{(i)}$.
Equation \rf{bbot} follows by a similar argument applied to the lower boundary.
The argument leading to \rf{ctmtop} requires the introduction
of the CTM of the homogeneous lattice \cite{JMN} denoted by
 $A_{{NW}}^{(i)}(\et)$ with one less argument. This is the CTM of
the same vertex model defined on the homogeneous
lattice of Figure 6.
\bfigt\vspace*{-15mm}\epsfysize=205mm \epsfbox{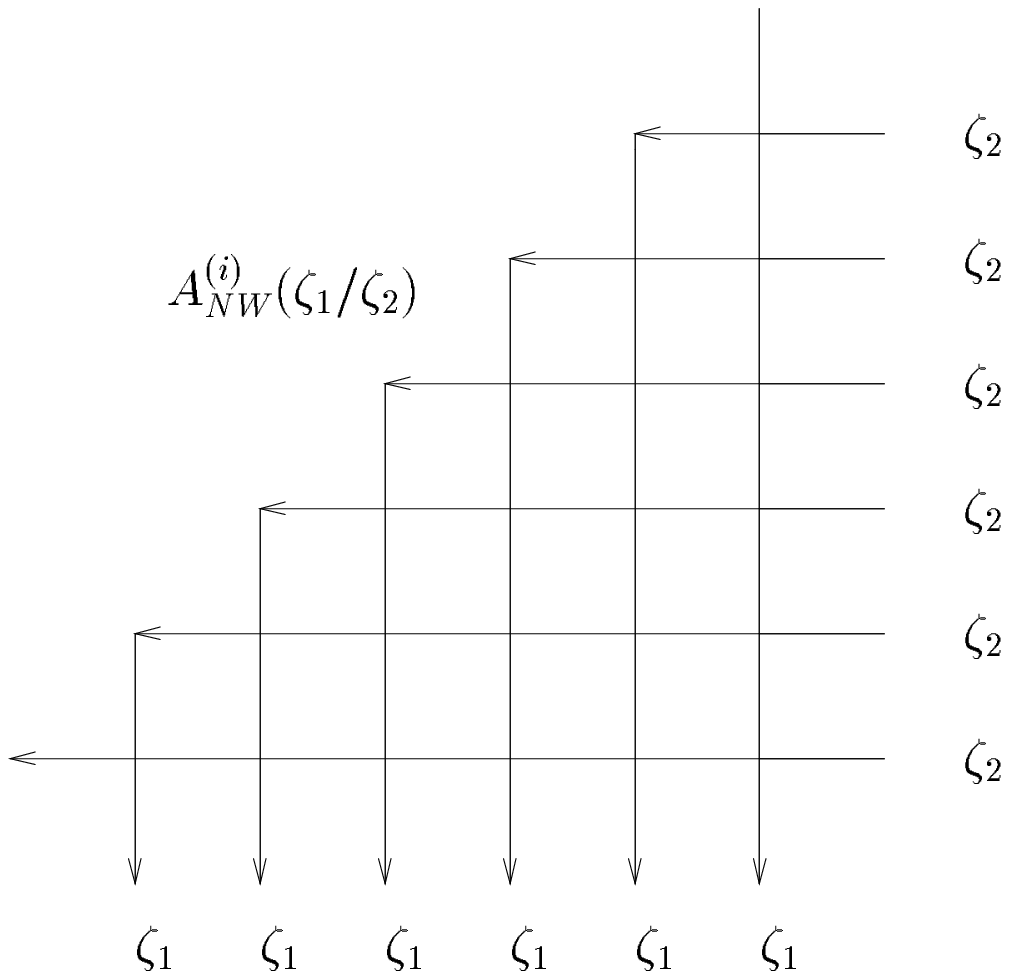}\vspace*{-115mm}%
\caption{The homogeneous CTM $A_{NW}^{(i)}(\z_1/\z_2)$}\efig
In \cite{JMN} a simple graphical argument is given that
leads to the equality
\be A_{NW}^{(i)}(\et) \p_{\vep}(\et) = \p_{\vep}(1) A_{NW}^{(1-i)}(\et)
\label{homo1}.\ee
This argument extends without change to the inhomogeneous lattice
discussed here, giving
\be A_{NW}^{(i)}(\z,\et) \p_{\vep}^U(\z,\et) = \p_{\vep}(1)
A_{NW}^{(1-i)}(\z,\et)
\label{inhomo1}.\ee
Baxter has shown that the asymptotic (i.e. infinite lattice) limit
of the CTM of the homogeneous 8-vertex model
is given by $A_{NW}^{(i)}(\et) = f(\et) \et^{D^{(i)}}$, where $D^{(i)}$ is
the CTM Hamiltonian (which is independent of $\et$) and
$f (\et)$ is a scalar function. It is then
a consequence of \rf{homo1} that
\be A_{NW}^{(i)}(\et) \p_{\vep}(\etp)
= \p_{\vep}(\etp/\et) A_{NW}^{(1-i)}(\et)
\label{homo2}.\ee
Finally, given \rf{homo2} and \rf{inhomo1}, property \rf{ctmtop}
follows if
\be A_{NW}^{(i)}(\et) A_{NW}^{(i)}(\z,\etp)
= g (\z,\et,\etp) A_{NW}^{(i)}(\z,\et \etp)
\label{key},\ee
where $g (\z,\et,\etp)$ is some scalar function.
In Appendix A we present a derivation of equation \rf{key}
using a generalization of Baxter's argument \cite{Bax82}
for the asymptotic behavior of the CTM of the homogeneous
8-vertex model.
If we apply $A_{NW}^{(i)}(\z,\et)$ to both side of \rf{btop}, and then use
\rf{ctmtop}, we obtain
\[
\sl_{\vep'} K_{\vep}^{\vep'}(\etp) \p_{\vep'}(\etp/\et) A_{NW}^{(i)}
 (\z,\et)\ket{B}^{(i)} =
\nu^{(i)}(\etp)\,\p_{\vep}({1 \ov \et \etp}) A_{NW}^{(i)}(\z,\et)
\ket{B}^{(i)}.
\]
Similarly,
\[
\sl_{\vep'} {^{(i)}{\bra{B}}} A_{SW}^{(i)} (\z,\et)
\pt_{\vep'}({1 \ov \et \etp})
K_{\vep'}^{\vep}(\etp) =
\nu^{(i)}(\etp)\,{^{(i)}{\bra{B}}} A_{SW}^{(i)}(\z,\et)\pt_{\vep}(\etp/\et).
\]

Specializing to the case $\et=1$, and making the identifications
\bac
\Phi_{\vep}(\z)&\sim&\phi_{\vep}(\z), \\
\Phi_{\vep}^{*}(\z)&\sim &\pt_{\vep}(\z), \\
\ket{0}_B&\sim & A_{NW}^{(0)}(\z,1) \ket{B}^{(0)}\ws \rm{and}\\
{_B{\bra{0}}}&\sim&{^{(0)}{\bra{B}}} A_{SW}^{(0)}(\z,1),
\label{ident}\ea
 we obtain equations
\rf{RBS} and \rf{LBS} of Section 2.
The remaining  property \rf{CR}
comes from \rf{yb}.

Thus, the vertex operators $\p_{\vep}(\et)$ and $\pt_{\vep}(\et)$, and
the boundary states $A_{NW}^{(0)}(\z,1) \ket{B}^{(0)}$ and
${^{(0)}{\bra{B}}} A_{SW}^{(0)}(\z,1)$ are lattice realizations of the
corresponding objects  discussed in section 2.
We speculate that the states $\ket{B}^{(0)}$ and ${}^{(0)}\bra{B}$ will
correspond to initial and final times states in the sense of \cite{GZ}
but now in an {\it{axial}} quantization scheme.
\subsection{Correlation functions and difference equations}
As in
\cite{JMN}, correlation functions of local operators
of the inhomogeneous vertex model
are defined in terms of
\be
{^{(i)}{\bra{B}}} \p_{{\vep}_1}^D (\z,\et)
\cdots \p_{\vep_N}^D (\z,\et)
A_{SW}^{(i+N)}(\z,\et)A_{NW}^{(i+N)}(\z,\et)
\p_{\vep_N}^U(\z,\et)\cdots \p_{\vep_1}^U(\z,\et) \ket{B}^{(i)},
\label{corrsimp}
\ee
where the superscripts on the CTM's are understood as modulo 2.
{}From properties \rf{ctmtop} and \rf{ctmbot} we see that it
is possible to move all the vertex  operators
$\p_{\vep_i}^U(\z,\et)$ and
$\p_{\vep_i}^D(\z,\et)$ inside of the
$A_{SW}^{(i)}(\z,\et) A_{NW}^{(i)}(\z,\et)$ product.
We can thus consider the more general spectral
parameter dependent
expression
\be
G^{(i)}(\z,\et|\et_1,\cdots \et_N)_{\vep_1,\cdots,\vep_N} =
{^{(i)}{\bra{B}}}A_{SW}^{(i)}(\z,\et)
\p_{\vep_1}(\et_1) \cdots \p_{\vep_N}(\et_N)
A_{NW}^{(i)}(\z,\et)
\ket{B}^{(i)}, \ws \rm{with}\ws N\ws \rm{even}.
\label{corrfn}
\ee
{}From \rf{ident}
$G^{(0)}(\z,\et=1|\et_1,\cdots,\et_N)_{\vep_1,\cdots,\vep_N}$
is the lattice
realization of the corresponding correlation function
$G(\et_1,\cdots,\et_N)_{\vep_1,\cdots,\vep_N}$ of equation (\ref{eqn:Gn}).
$G^{(0)}(\z,\et=1|\et_1,\cdots,\et_N)_{\vep_1,\cdots,\vep_N}$
thus obeys the difference equation \rf{CHERDIFF}
from the arguments
already given in section 2.
Here we shall attempt to give some
insight into the physical origin of this difference equation
by pointing out the sequence of steps necessary in order to
derive it within the context of the lattice theory.
For simplicity, we drop the $\vep$ subscripts. At each step
we refer to the vertex operator that arises as a result of
the previous step.
\newcounter{list1}
\begin{list}%
{\arabic{list1}}{\usecounter{list1}%
\setlength{\leftmargin=5mm}{\itemsep=0mm}}
\item Move $\p(\et_j)$ to the right of
$\p(\et_{j+1}) \cdots \p(\et_N)$
using the exchange relation \rf{yb}.
\item Move $\p(\et_j)$ to the right of the CTM $A_{NW}(\z,1)$ using the
commutation property \rf{ctmtop}.
\item `Reflect' $\p^{U}(\z,\et_j )$ off the upper boundary using
\rf{btop}.
\item Move $\p^{U}(\z,1/\et_j)$ back to the left of
the CTM  $A_{NW}(\z,1)$ using \rf{ctmtop}.
\item Move $\p(1/\et_j)$ to the left of
$\p(\et_1) \cdots \p(\et_{j-1}) \p(\et_{j+1}) \cdots \p(\et_N)$
using \rf{yb}.
\item `Cross' $\p(1/\et_j)$ using the crossing relation \rf{cross}.
\item Move $\pt(-q/\et_j)$ to the left of $A_{SW}(\z,1)$ using
the commutation property \rf{ctmbot}.
\item Reflect $\p^D(\z,-q/\et_j)$ off the lower boundary using
\rf{bbot}.
\item Move $\p^D(\z,-\et_j/q)$ to the right of $A_{SW}(\z,1)$
using \rf{ctmbot}.
\item Cross $\pt(-\et_j/q)$ using \rf{cross}.
\item Move $\p(\et_j/q^2)$ to the right of $\p(\et_1) \cdots \p(\et_{j-1})$
using \rf{yb}.
\end{list}
This sequence of steps leads to the difference equation
\rf{CHERDIFF}. One can of course derive the same equation by moving
$\p(q^{-2}\et_j)$ to the left of $\p(\et_1) \cdots \p(\et_{j-1})$,
through $A_{SW}(\z,1) $, reflecting off the lower boundary, etc.
\setcounter{equation}{0}
\section{Boundary spontaneous magnetization}
In this section, we investigate the difference equation obeyed by the two
point function of the $\XYZ$ spin chain.  The solution of this
equation, with a certain
specialization of the spectral parameters, gives the boundary
spontaneous magnetization.  For general values of the boundary magnetic field,
the
solution does not factorize, but in the case of free boundary
conditions, we find a factorized solution. As we show below, the value
of the boundary spontaneous magnetization is (minus) the square of
the spontaneous staggered polarization of the usual eight-vertex model
without
boundary.

\subsection{Two point functions}
Setting $n=2$ in the elliptic version of \rf{n-point}, we have
\bea
&&G(q^{-2}\z_1,\z_2)=\overline{K}_1(\z_1)
R_{21}(\z_1\z_2)K_1(\z_1)R_{12}(\z_1/\z_2)
G(\z_1,\z_2), \label{tptdiffa}\\
&&G(\z_1,q^{-2}\z_2)=R_{21}(q^{-2}\z_2/\z_1)\overline{K}_2(\z_2)
R_{12}(\z_1\z_2)K_2(\z_2)G(\z_1,\z_2), \label{tptdiffb}
\ena
where
\bea
&&G(\z_1,\z_2)=\sum_{\vep_1,\vep_2}G_{\vep_1,\vep_2}(\z_1,\z_2)
v^{\vep_1}\otimes v^{\vep_2},\nonumber\\
&&G_{\vep_1,\vep_2}(\z_1,\z_2)={}_B\bra{i}\Phi_{\vep_1}(\z_1)
\Phi_{\vep_2}(\z_2)\ket{i}_B .\nonumber
\ena

In the following paragraph, we restrict our attention to the free
boundary condition, $r=-1$, or $h=0$.  The reflection matrices $K$ and
$\overline{K}$ in this limit are
$$
K(\z)=\frac{1}{f(\z)}I, \qquad \overline{K}(\z)=\frac{1}{f(-q^{-1}\z)}I.
$$
Here $I$ denotes the 2$\times$2 unit matrix and $f(\z)=f(\z;-1)$.

Let
\be
G_{\pm}(\z_1,\z_2)=G_{+-}(\z_1,\z_2)\pm G_{-+}(\z_1,\z_2).\lb{Gpm}
\en
One can reduce equations (\ref{tptdiffa}), (\ref{tptdiffb}) to the
following scalar difference equations:
\bea
G_{\vep}(q^{-2}\z_1,\z_2)=
\frac{\nu({\vep}\z_1\z_2)\nu({\vep}\z_1/\z_2)}{f(\z_1)f(-q^{-1}\z_1)}
G_{\vep}(\z_1,\z_2) \label{sdiffa},\\
G_{\vep}(\z_1,q^{-2}\z_2)=
\frac{\nu({\vep}q^{-2}\z_2/\z_1)\nu({\vep}\z_1\z_2)}{f(\z_2)f(-q^{-1}\z_2)}
G_{\vep}(\z_1,\z_2), \label{sdiffb}
\ena
where
$$
\nu(\z)=\frac{1}{\overline{\kappa}(\z^2)}
\frac{(-q\z^{-1};p)_{\infty} (-pq^{-1}\z;p)_{\infty}}{
(-q\z;p)_{\infty} (-pq^{-1}\z^{-1};p)_{\infty}},
$$
and $\overline{\kappa}(z)$ is defined in \rf{kappabar}.
We solve these equations by setting
$$
G_\vep(\z_1,\z_2)=A(\z_1)A(\z_2)B_\vep(\z_1\z_2)B_\vep(\z_1/\z_2).
$$
The problem is now reduced to the following equations:
\bean
&&\frac{A(q^{-2}\z)}{A(\z)}=\frac{1}{f(\z)f(-q^{-1}\z)}, \cr
&&\frac{B_\vep(q^{-2}\z)}{B_{\vep}(\z)}=\vep\ \nu({\vep}\z).
\enan
The solutions which are analytic in the region $-q<|\z|<q^{-2}$  are given by
\bean
&&A(\z)=\frac{(q^2z^{-1};p,q^4)_{\infty}
(q^6z^{-2};p,q^4,q^8)_{\infty}
(pq^2z^{-2};p,q^4,q^8)_{\infty}}{
(pz^{-1};p,q^4)_{\infty}
(q^4z^{-2};p,q^4,q^8)_{\infty}
(pq^4z^{-2};p,q^4,q^8)_{\infty}}\cr
&&\qquad\qquad\times
\frac{(q^4z^{};p,q^4)_{\infty}
(q^{10}z^{2};p,q^4,q^8)_{\infty}
(pq^6z^{2};p,q^4,q^8)_{\infty}}{
(pq^2z^{};p,q^4)_{\infty}
(q^8z^{2};p,q^4,q^8)_{\infty}
(pq^8z^{2};p,q^4,q^8)_{\infty}},\label{sola}\\
&&B_+(\z)=
\frac{(-pq^{-1}\z^{-1};p,q^2)_{\infty}
(q^2z^{-1};p,q^4,q^4)_{\infty}
(pq^2z^{-1};p,q^4,q^4)_{\infty}}{
(-q\z^{-1};p,q^2)_{\infty}
(q^4z^{-1};p,q^4,q^4)_{\infty}
(pz^{-1};p,q^4,q^4)_{\infty}}\cr
&&\qquad\qquad\times\frac{(-pq^{}\z^{};p,q^2)_{\infty}
(q^6z^{};p,q^4,q^4)_{\infty}
(pq^6z^{};p,q^4,q^4)_{\infty}}{
(-q^3\z^{};p,q^2)_{\infty}
(q^8z^{};p,q^4,q^4)_{\infty}
(pq^4z^{};p,q^4,q^4)_{\infty}},\\
&&B_-(\z)=B_+(-\z),
\enan
where $z=\z^2$.

\subsection{Spontaneous magnetization}
The boundary magnetization is the vacuum expectation value of the
boundary spin operator $\sigma_1^z$. As in the $\XXZ$ chain, it is
given by the value at $\z=1$ of the following function:
$$
{\cal M}^{(0)}(\z;r)=\frac{{}_B\bra{0}E^{(0)}_{++}(\z,\z)-
E^{(0)}_{--}(\z,\z)\ket{0}_B
}{{}_B\langle {0}| {0}\rangle_B},\label{magnetz}
$$
where
$$
E^{(i)}_{\vep,\vep'}(\z_1,\z_2)=\Phi^{*(i,1-i)}_\vep(\z_1)
\Phi^{(1-i,i)}_{\vep'}(\z_2).
$$
In terms of the functions $G_\vep(\z_1,\z_2)$ of (\ref{Gpm}),
\be
{\cal M}^{(0)}(\z;r)=-\frac{G_-(-q^{-1}\z, \z)}{G_+(-q^{-1}\z,\z)}.
\en
Using the result in the last section for the case $r=-1$, we obtain
\bean
&&{\cal M}^{(0)}(\z;-1)=-
\frac{(-pz^{};p,q^2)_{\infty}
(-pz^{-1};p,q^2)_{\infty}
(-p;p,q^2)^2_{\infty}}{
(pz^{};p,q^2)_{\infty}
(pz^{-1};p,q^2)_{\infty}
(p;p,q^2)^2_{\infty}}\cr
&&\qquad\qquad\times\frac{(q^2 z^{};p,q^2)_{\infty}
(q^2z^{-1};p,q^2)_{\infty}
(q^2;p,q^2)^2_{\infty}}{
(-q^2 z^{};p,q^2)_{\infty}
(-q^2z^{-1};p,q^2)_{\infty}
(-q^2;p,q^2)^2_{\infty}}.
\enan
The value at $\z=1$ gives the spontaneous magnetization of the
boundary spin operator:
\be
{\cal M}^{(0)}(1;-1)=-\frac{
(q^2;q^2)^4_{\infty}(-p;p)^4_{\infty}}
{(-q^2;q^2)^4_{\infty}(p;p)^4_{\infty}}. \label{spomag}
\en
The $\XXZ$ limit $p\to 0$ coincides with the previous result of
\cite{JKKKM}.  We again note the remarkable fact that the
spontaneous magnetization $-{\cal M}^{(0)}(1;-1)$ is exactly the
square of the corresponding bulk quantity, i.e. the spontaneous
staggered polarization in the eight-vertex model\cite{BaKe,JMN}. The
same phenomenon was observed in the case of the $\XXZ$ chain
in~\cite{JKKKM}.

One can check formula (\ref{spomag}) by comparing it with the
derivative with respect to the external magnetic field $h$ of the
energy difference $\Delta e(r)=e^{(1)}(r)-e^{(0)}(r)$.
{}From (\ref{energy}), we obtain
$$
\frac{\partial \Delta e(r)}{\partial h}
=2\frac{
(q^2;q^2)^4_{\infty}
(r^2q^2;q^4)_{\infty}
(r^{-2}q^2;q^4)_{\infty}}{(q^2;q^4)^2_{\infty}
(rq^2;q^2)^2_{\infty}
(r^{-1}q^2;q^2)^2_{\infty}}
\frac{(p;p^2)^2_{\infty}(rp;p)^2_{\infty}(r^{-1}p;p)^2_{\infty}}
{(p;p)^4_{\infty}
(r^2p;p^2)_{\infty}
(r^{-2}p;p^2)_{\infty}}.
$$
This quantity is equal to the difference of the magnetizations
${\cal M}^{(1)}(1;r)-{\cal M}^{(0)}(1;r)$, which agrees, at $r=-1$,
with the result \rf{spomag}.

One can also verify the result for $\cM^{(0)}(1;-1)$ directly using
perturbation theory in $\vep={-q / (1+q^2)}$. As in \cite{JKKKM}, we
solve order by order  the equation
$$
\left(\sl_{k \ge 1}\left( \half (\s_{k+1}^z \s_k^z +1) +c_k(\vep)
\right) -2 \vep (Q+\G Q^{\prime})\right) |0\rangle_B =0,
$$
 where
$$ Q=\sl_{k \ge 1}(\s_{k+1}^+ \s_{k}^- + \s_{k+1}^-
\s_{k}^+), \qquad Q^{\prime}=\sl_{k \ge 1}(\s_{k+1}^+ \s_{k}^+ +
\s_{k+1}^- \s_{k}^-).
$$
 The c-number normalization term
$c_k(\vep) =\sum_{j \ge 1} c_{k,j}\vep^j$ is included in order to
ensure that the eigenvalue is zero.  Solving for $\ket{0}_B$, we find
$$ \cM^{(0)}(1;-1)= -1 + 8 \vep^2 + 8(5 \G^2 -1) \vep^4 +O(\vep^6). $$
In terms of $q$ and $p$, this expansion becomes
$$ \cM^{(0)}(1;-1)= -1 + 8 q^2 -24 q^4 -8 p + 0(q^6,p q^2,p^2),$$
which agrees with the expression \rf{spomag}.
\setcounter{equation}{0}
\setcounter{section}{4}
\section{Boundary $S$ matrix}
We now come to a brief discussion about the $S$-matrices which describe
scatterings of quasi-particles with themselves or with the boundary.
These $S$-matrices obey the same type of Yang-Baxter equations,
unitarity and crossing relations, as do the $R$- and the $K$-matrices
used to construct the model, but they are not the same.
In fact, for the $\XYZ$ chain in the bulk, one expects \cite{FIJKMY}
that the two-particle $S$-matrix has the same form
as $R$ but with the elliptic parameter $p$ being changed:
\begin{equation}\label{S}
S(\xi;p,q)=-R(\xi;p^*,q),
\qquad p^*=pq^{-2}.
\end{equation}
A similar phenomenom was observed in \cite{JKKKM}, where the
the boundary $S$-matrix $M(\xi)$ for the $\XXZ$ chain was shown to be of
the same form as the $K$-matrix, up to an overall scalar factor,
wherein the parameter $r$ is changed to $r^*=rq^{-1}$.

We are then led to
speculate that the $M$-matrix for the $\XYZ$ chain with a boundary
(see \rf{b-s} for the characterization) is proportional to the $K$-matrix,
with both $p$ and $r$ scaled in the same way as above; namely
\begin{equation}\label{M}
M(\xi;p,q,r)=
\frac{1}{\overline{f}(\xi;p,q,r)}
\widehat{K}(\xi;p^*,r^*),
\qquad p^*=pq^{-2},~ r^*=rq^{-1},
\end{equation}
where by $\widehat{K}(\xi;p^*,r^*)$ we mean
the matrix $\widehat{K}(\zeta;p,r)=\widehat{K}(\zeta;r)$ in (\ref{KMAT})
with the parameters $p,r$ being replaced by $p^*,r^*$ respectively.
The scalar factor $\overline{f}(\xi;p,q,r)$ is fixed by solving the
boundary unitarity and crossing relations,
\begin{eqnarray*}
&&M(\xi)M(\xi^{-1})=1,
\\
&&M_a^b(-q^{-1}\xi^{-1})=
\sum_{a',b'}S^{b'\,-a'}_{-b\,a}(-q\xi^2)M^{a'}_{b'}(\xi).
\end{eqnarray*}
with the condition that
it reduces to the known result \cite{JKKKM}
in the $\XXZ$ limit $p\rightarrow 0$.
Explicitly we have
\[
\frac{1}{\overline{f}(\xi;p,q,r)}=
-\xi^{-2}
\frac{\phib(\xi^2;p^*,q,r^*)}{\phib(\xi^{-2};p^*,q,r^*)},
\]
where
\begin{eqnarray*}
\phib(z;p^*,q,r^*)&=&\phib_0(z;p^*,q)\phib_1(z;p^*,q,r^*),
\\
\phib_0(z;p^*,q)&=&
\frac{(q^2z^2;p^*,q^8)_\infty}{(q^8z^2;p^*,q^8)_\infty}
\frac{(p^*q^6z^2;p^*,q^8)_\infty}{(p^*z^2;p^*,q^8)_\infty},
\\
\phib_1(z;p^*,q,r^*)&=&
\frac{(p^*q^8z^2;p^{*2},q^8)_\infty(p^*z^2;p^{*2},q^8)_\infty}
{(p^*q^4z^2;p^{*2},q^8)^2_\infty}
\\
&&
\times
\frac{(q^4r^*z;p^*,q^4)_\infty}{(q^2r^*z;p^*,q^4)_\infty}
\frac{(p^*r^{*-1}z;p^*,q^4)_\infty}{(p^*q^2r^{*-1}z;p^*,q^4)_\infty}
\frac{(p^*r^*z;p^{*2})_\infty}{(p^*r^{*-1}z;p^{*2})_\infty}.
\end{eqnarray*}

In the case of the $\XXZ$ chain, it was possible to derive the $M$-matrix
since the boundary vacuum states are known explicitly in terms of
the bosonic oscillators.
Here we do not have such a construction, and
(\ref{M}) is no more than a plausible guess.
We find it difficult to check it by perturbative methods.
As an alternative argument in its favor, we
study below the continuum limit
and show that the resulting formulas agree
with those for the sine-Gordon model with a boundary \cite{GZ}.

Introduce
\[
\hat{\lambda}=\frac{2\lambda}{\pi-2\lambda}
\]
and write
\begin{equation}
\xi^2=(p^*)^{-\tilde{\lambda}u/\pi},
\quad
q^2=(p^*)^{\tilde{\lambda}},
\quad
r^*=(p^*)^{1/2-\hat{\eta}/\pi}.
\end{equation}
By the continuum limit we mean the limit
$K\rightarrow \infty$, $K'\rightarrow \pi/2$,
so that $p^*\rightarrow 1$,  $\tilde{\lambda}\rightarrow \hat{\lambda}$,
while keeping $u$ and $\hat{\eta}$ fixed.
We find the following:
\begin{eqnarray}
\lim S(\xi)&=&\rho(u)\times
\Bigl(
\sin\bigl(\hat{\lambda}\pi\bigr)
{}~\frac{1}{2}\bigl(1+\sigma^y\otimes\sigma^y)
+
\sin\bigl(\hat{\lambda}(\pi-u)\bigr)
{}~\frac{1}{2}\bigl(\sigma^x\otimes\sigma^x+\sigma^z\otimes\sigma^z)
\nonumber\\
&&
+
\sin\bigl(\hat{\lambda}u\bigr)
{}~\frac{1}{2}\bigl(\sigma^x\otimes\sigma^x-\sigma^z\otimes\sigma^z)
\Bigr),
\label{limS}\\
\lim M(\xi)&=&
-R_0(u)\overline{\sigma}(0,u)\overline{\sigma}(\hat{\eta},u)
\cos(\hat{\lambda}u)
\times\left(1-\frac{\sin(\hat{\lambda}u)}{\cos\hat{\eta}}\sigma^z\right).
\label{limM}
\end{eqnarray}
Here the functions $\rho(u)$, $R_0(u)$, $\sigma(x,u)$ are
given in (5.7), (5.21), (5.23) in \cite{GZ}, respectively,
and $\overline{\sigma}(x,u)=\sigma(x,u)/\sigma(x,0)$, wherein
our $\hat{\lambda}$ and $\hat{\eta}$ are to be
identified with $\lambda$ and $\eta$ there.
\footnote{It seems that in (5.22), \cite{GZ}, one should use
$\overline{\sigma}(x,u)=\sigma(x,u)/\sigma(x,0)$ in place of
$\sigma(x,u)$, to ensure $R_1(0)=1/\cos \xi$.}

The formula (\ref{limS}) differs from the standard formula
(5.6-7) \cite{GZ} for the two-particle $S$-matrix of the sine-Gordon theory
by a gauge transformation.
To see this, let
\[
U=\pmatrix{1&-i\cr 1& i\cr},
\qquad
\tilde{U}=\sigma^z\, U \,\sigma^z,
\]
which has the property $U\sigma^xU^{-1}=\sigma^y$,
$U\sigma^yU^{-1}=\sigma^z$, $U\sigma^zU^{-1}=\sigma^x$.
Let further
\[
\psi^{*(j)}_\mu(\xi)=\sum_{\nu}U_\mu^{(j)\nu}\Psi^*_\nu(\xi),
\]
where $U^{(j)}=U$ for $j$ odd and
$=\tilde{U}$ for $j$ even.
Define the new basis of eigenstates by
\begin{equation}
\ket{\xi_n,\cdots,\xi_1}_{\mu_n,\cdots,\mu_1;(i)B}'
=
\psi^{*(n)}_{\mu_n}(\xi_n)\cdots
\psi^{*(2)}_{\mu_2}(\xi_2)\psi^{*(1)}_{\mu_1}(\xi_1)
\ket{i}_{B}.
\end{equation}
In this basis the bulk and the boundary $S$-matrices are given by
\[
S'(\xi)=\left(\tilde{U}\otimes U\right)
S(\xi)\left(U\otimes \tilde{U}\right)^{-1},
\qquad
M'(\xi)=U\, M(\xi)\, U^{-1}.
\]
Then $\lim\,S'(\xi)$ coincides with the formula (5.6-7) of \cite{GZ}.
The limit of the $M'$-matrix is to be compared with a special case of
the boundary $S$-matrix in \cite{GZ}
\[
\hat{\xi}=0, \quad \vartheta=0,
\]
where $\hat{\xi}$ denotes the parameter $\xi$ in \cite{GZ}.
With this specialization the $\lim\,M'(\xi)$ agrees with (5.12), (5.21--25)
in \cite{GZ} up to an overall sign.
In particular, the free boundary condition (5.29) \cite{GZ} of the
sine-Gordon theory is given by
$\hat{\eta}=\pi(\hat{\lambda}+1)/2$.
It corresponds exactly to the free boundary condition
$r=-1$ in the $\XYZ$ chain.
This is an indication that our speculative choice of the parameter
$r$ in the $M$-matrix (\ref{M}) is exact.
\setcounter{equation}{0}
\setcounter{equation}{0}
\setcounter{section}{5}
\section{Properties of boundary form factors}
We now consider the boundary form factors, defined as
\be
F^{(i)}_n(\xi_1,\ldots,\xi_n) = \sum_{\vep_1,\ldots,\vep_n}
v^*_{\vep_1}\otimes\cdots\otimes v^*_{\vep_n}~
{}_B\langle i | {\cal O} \Psi^*_{\vep_1}(\xi_1) \cdots \Psi^*_{\vep_n}(\xi_n)
|i\rangle_B~ ,\label{ff-def}
\en
where ${\cal O}$ is some local operator, which commutes with type II
vertex operators. Here, $v^*_\vep$ denote basis vectors in the space
dual to $V$.  The boundary form factor satisfies properties analogous
to Smirnov's axioms for form factors, but with some differences due to
the presence of a boundary.  These properties follow from the
properties of type II vertex operators and the boundary vacuum states.
The commutation relations for type II vertex operators for the $\XYZ$
model were presented in~\cite{FIJKMY}. In particular, $$
\Psi^*_{\vep_1}(\xi_1)\Psi^*_{\vep_2}(\xi_2) =\sum_{\vep_1',\vep_2'}
\Psi^*_{\vep_2'}(\xi_2)\Psi^*_{\vep_1'}(\xi_1)
S_{\vep_1,\vep_2}^{\vep_1',\vep_2'}(\xi_1/\xi_2)~,
$$
where the $S$-matrix is defined in \rf{S}.

By definition of the boundary $S$-matrix, type II vertex operators
have the following property with respect to the boundary bound states:
\begin{eqnarray}
\Psi^*_\vep(\xi) |i\rangle_B = \sum_{\vep'}
{M^{(i)}}^{\vep'}_\vep(\xi) \Psi^*_{\vep'}(\xi^{-1})|i\rangle_B, \nonumber\\
{}_B\langle i|\Psi^*_\vep(\xi^{-1})= \sum_{\vep'}{}_B\langle
i|\Psi^*_{\vep'}(q^{-2}\xi) {\overline{M}^{(i)}}{}^{\vep'}
_\vep(\xi), \label{b-s}
\end{eqnarray}
where the boundary $S$-matrices $\overline{M}^{(i)}(\xi)
=\sigma^x{M}^{(i)}(-q^{-1}\xi)\sigma^x$, $M^{(0)}(\xi)=M(\xi)$ of Eq.
\rf{M}, and $M^{(1)}(\xi;r) = \sigma^x M^{(0)}(\xi;r^{-1})\sigma^x$.

Let
\begin{eqnarray*}
(\cdots\otimes v^*_{\vep_j}
\otimes v^*_{\vep_i}\otimes\cdots)~
S_{i,j}(\xi) &=& \sum_{\vep_i',\vep_j'}
(\cdots\otimes v^*_{\vep_j'}
\otimes v^*_{\vep_i'}\otimes\cdots)~
S_{\vep_i,\vep_j}^{\vep_i',\vep_j'}(\xi)~,\\
(\cdots\otimes v^*_{\vep_{i+1}}\otimes v^*_{\vep_i}\otimes\cdots)~P_{i,i+1}
&=&
\cdots\otimes v^*_{\vep_i}\otimes v^*_{\vep_{i+1}}\otimes\cdots~,\\
(\cdots\otimes v^*_{\vep_j}\otimes\cdots)~ M^{(i)}_j(\xi) &=&
\sum_{\vep_j'}(\cdots\otimes v^*_{\vep_j'}\otimes\cdots) ~
{M^{(i)}}^{\vep_j}_{\vep_j'}(\xi)~ ,
\end{eqnarray*}
etc..
Then the boundary form factors defined in (\ref{ff-def})
have the following properties, analogous to Smirnov's form factor
axioms.

\noindent{\bf Axiom I:}
Due to the exchange relation between type II vertex operators, which
remains unchanged in the boundary theory, this property is identical
to the usual case in the absence of a boundary:
$$
F_n^{(i)}(\xi_1,\ldots,\xi_j,\xi_{j+1},\ldots\xi_n) =
F_n^{(i)}(\xi_1,\ldots,\xi_{j+1},\xi_j,\ldots\xi_n)
 P_{j,j+1}{S}_{j,j+1}(\xi_j/\xi_{j+1})~.
$$

\noindent{\bf Axiom II:} The analog of the periodicity condition is
the difference equation satisfied by the form factors, which is
similar to Cherednik's equation \rf{CHERDIFF}. It is a consequence of
Axiom I and the properties (\ref{b-s})
\begin{eqnarray*}
F_n^{(i)}(\xi_1,\ldots,q^{2}\xi_j,\ldots,\xi_n) &=&
F_n^{(i)}(\xi_1,\ldots,\xi_j,\ldots,\xi_{n})
S_{j,j-1}(\xi_j/\xi_{j-1})\cdots S_{j,1}(\xi_j/\xi_1)
\overline{M}^{(i)}_j(q^2\xi_j)
\\ &&\hbox{\hskip-.6in}\times~
S_{1,j}(q^2\xi_1\xi_j)\cdots S_{j-1,j}(q^2\xi_{j-1}\xi_j)
S_{j+1,j}(q^2\xi_{j+1}\xi_j)\cdots S_{n,j}(q^2\xi_n\xi_j)
M^{(i)}_j(q^2\xi_j)
\\ &&\hbox{\hskip-.6in}\times~
S_{j,n}(q^2\xi_j/\xi_n)\cdots S_{j,j+1}(q^2\xi_j/\xi_{j+1})~.
\end{eqnarray*}

\noindent{\bf Axiom III:} The form factor (\ref{ff-def}) has simple
annihilation poles due to the relation (see equation (B.9) in
\cite{FIJKMY2}\footnote{There is an error in equation (B.9): The first
factor is the inverse of the correct expression.})
$$
\Psi^{*(i,1-i)}_{\vep_1}(\xi_1) \Psi^{*(1-i,i)}_{\vep_2}(\xi_2) =
\frac{g^* ~\delta_{\vep_1,-\vep_2}}{1-q^{-2} \xi_2^2/\xi_1^2} \left(
-\frac{\xi_2} {q\xi_1} \right)^{i+{1+\e_1\over 2}} + \cdots~,~~~\xi_1
\rightarrow \pm q^{-1} \xi_2~,
$$
where ``$\cdots$'' refers to regular terms and the scalar $g^*$ is as in
Eq. \rf{gxyz} with $p$ replaced by $p^*=p/q^2$.
The residue of the form factor at the point $\xi_j =-q^{-1}\xi_n$ is
\begin{eqnarray}
{\rm Res}F_n^{(i)}
(\xi_1,\ldots,\xi_j,\ldots,\xi_n) d(\xi_j/\xi_n)&=& \frac{g}{2}
F_{n-2}^{(i)}(\xi_1,\ldots,\xi_{j-1},\xi_{j+1},\ldots,\xi_{n-1})\otimes C
\nonumber\\
&&\hbox{\hskip-2.1in} \times P_{n-2,n-1}\cdots P_{j,j+1}~
\Bigg( S_{j,n-1}(-\xi_{n}/q\xi_{n-1})\cdots S_{j,j+1}(-\xi_{n}/q\xi_{j+1})
\nonumber\\
&& \hbox{\hskip-2.1in} -
M_n(\xi_n)
S_{j,n-1}(-1/q\xi_{n-1}\xi_n)\cdots S_{j,j+1}(-1/q\xi_{j+1}\xi_n)
S_{j,j-1}(-1/q\xi_{j-1}\xi_n)
\nonumber\\ && \hbox{\hskip-2.1in}\times \cdots S_{j,1}(-1/q\xi_1\xi_n)
\overline{M}^{(i)}_j(-q\xi_n^{-1}) S_{1,j}(-q\xi_1/\xi_n)\cdots S_{j-1,j}
(-q\xi_{j-1}/\xi_{n})
\Bigg)~,\lb{axiomIII}
\end{eqnarray}
where $C=\sum_{\vep_1,\vep_2}
v^*_{\vep}\otimes v^*_{-\vep}$. There are
additional poles corresponding to bound states, which come from the
poles in the $S$-matrix and boundary $S$-matrix.
\setcounter{equation}{0}
\section{Discussion}
Let us summarize our results.  We have shown that correlation
functions of the semi-infinite $\XXZ$ spin chain obey the quantum
Knizhnik--Zamolodchikov equation with reflection.  This is a simple
consequence of the exchange algebra of the vertex operators and the
reflection property \rf{RBS},\rf{LBS} of the boundary vacuum state
and its dual.

In the case of the $\XYZ$ chain, we assumed such a boundary vacuum
state exists with a similar reflection property. Again using the
exchange algebra between the elliptic vertex operators of
\cite{FIJKMY} and the reflection property with the elliptic $K$-matrix
of \cite{IK,HOU} we were able to derive the $q$-difference equations
satisfied by the correlation functions of the $\XYZ$ spin chain with a
boundary. We cannot obtain an explicit formula for the vacuum states
$|i\rangle_B$ in this case, however, due to a lack of a bosonization
formula for the elliptic vertex operators.

The assumptions made in the latter case are supported by an alternative
method of derivation using the corner transfer matrix of the eight
vertex model with a boundary corresponding to Sklyanin's transfer
matrix.  By constructing a lattice realization of type I vertex
operators and boundary vacuum states, we showed that they obey the
appropriate commutation relations and reflection condition, and
therefore that the correlation functions obey the correct difference
equations.  Using the difference equations, we computed the boundary
spontaneous magnetization ${\cal M}^{(0)}(1;-1)$, i.e., the
magnetization in the case where the boundary field vanishes.

We also conjectured a natural form for the boundary $S$-matrix for the
$\XYZ$ model, which is the reflection matrix associated with type II
vertex operators.  In the continuum limit the boundary $S$-matrix
becomes (up to a gauge transformation) that of the boundary
sine-Gordon model~\cite{GZ}.  Finally, using this boundary $S$-matrix
and the exchange algebra of the elliptic type II vertex operators, we
formulated the analogues of Smirnov's axioms for form factors
\cite{Smi92}.

For the $\XYZ$ chain, neither the bulk nor the boundary
$S$-matrix have been derived using an alternative method to the vertex
operator approach. It should be possible to obtain these using the
Bethe ansatz approach. The same can be said for the boundary
$S$-matrix of the $\XXZ$ chain.

In our boundary $S$-matrix, one can find a pole $\xi^2=rq^{-1}$ in the
physical strip $1<|\xi^2|<|q|^{-2}$. This pole yields a boundary bound
state, and as in the $\XXZ$ chain case~\cite{JKKKM} this state should be
identified with the second vacuum state $\ket{1}_B$. The coincidence
of the vanishing point $r=q, -q$ of ${\partial
\Delta e}/{\partial h}$ and the point where the pole leaves the
physical strip gives a consistency check for
the conjectured form of the boundary $S$-matrix.

In section 6, we derived the difference equation for form
factors of the semi-infinite spin chain. In the continuum limit,
one can derive the difference equations satisfied by form factors in
massive integrable quantum field theory on a semi-infinite line.
For example, for the sine-Gordon theory, using the setting given in
Section 5 and taking the limit $p^*\to 1 $ of the equation \rf{axiomIII},
one finds
\bea
&&{\cal F}_n(\beta_1,\cdots,\beta_j+2\pi i, \cdots, \beta_n )\cr
&&={\cal F}_n(\beta_1,\cdots,\beta_j, \cdots, \beta_n )
{\cal S}_{jj-1}(\beta_{j}-\beta_{j-1})\cdots
{\cal S}_{j1}(\beta_{j}-\beta_1)
{\bar{\cal M}}_j(\beta_j-2\pi i)
\cr
&&\quad\times{\cal S}_{1j}(\beta_{j}+\beta_1-2\pi i)\cdots
{\cal S}_{j-1j}(\beta_j+\beta_{j+1}-2\pi i)\cr
&&\quad\times{\cal S}_{j+1j}(\beta_j+\beta_{j-1}-2\pi i)\cdots
{\cal S}_{nj}(\beta_j+\beta_{1}-2\pi i)
{{\cal M}}_j(\beta_j-2\pi i)
\cr
&&\quad\times
{\cal S}_{jn}(\beta_{j}-\beta_n-2\pi i)\cdots
{\cal S}_{jj+1}(\beta_j-\beta_{j+1}-2\pi i)
,
\ena
where we also set $u=-i\beta$ and
\bea
&&\lim {F_n'}(\xi_1, \cdots, \xi_n )={\cal F}_n(\beta_1,\cdots, \beta_n
), \cr
&&\lim S'(\xi)={\cal S}(\beta),\quad \lim M'(\xi)={\cal M}(\beta).
\ena
Here the primed quantities are the gauge transformed ones (see Sec.5).
To solve this equation is an open problem. It seems possible by
applying the method developed by Smirnov\cite{Smi92} with modification by
Sklyanin's Bethe ansatz scheme\cite{Skl87}.

Taking the quasi-classical limit of our difference equation, one can
derive the boundary analog of the KZ equation. It is however unclear
whether this equation governs the boundary conformal field theory in
the way that the KZ equation does for the bulk theory. It would
be nice if one could relate the equation to the one obtained by
Cardy\cite{Car}.

\vskip 1cm
\noindent{{\Large Acknowledgements.}\quad}
We wish to thank
D. Bernard,
P. Fendley,
T. Inami
and
A. Ludwig
for useful discussions.
R. A. W would like to thank P. Dorey for initially
suggesting the CTM approach to boundary problems, and his colleagues
at RIMS for their hospitality during the period in which much of this
work was carried out.
This work is partly supported by Grant-in-Aid for Scientific Research
on Priority Areas 231, the Ministry of Education, Science and Culture.
H. K. is supported by Soryushi Shyogakukai.
R. K. is supported by the Japan Society for the Promotion of Science.
\baselineskip=17pt
\appendix
\setcounter{figure}{0}
\renewcommand{\thefigure}{\thesection.\arabic{figure}}
\setcounter{equation}{0}
\newpage
\section{The Asymptotic Behaviour of the CTM}
In this appendix we derive the asymptotic form of the
CTM of the inhomogeneous 8-vertex model using a
generalisation of Baxter's argument for the homogeneous
model.
Consider the transfer matrix $T(\{\z_i\},\{\et_i\})$ shown
in Figure A1.

\bfig[here]\epsfysize=180mm \epsfbox{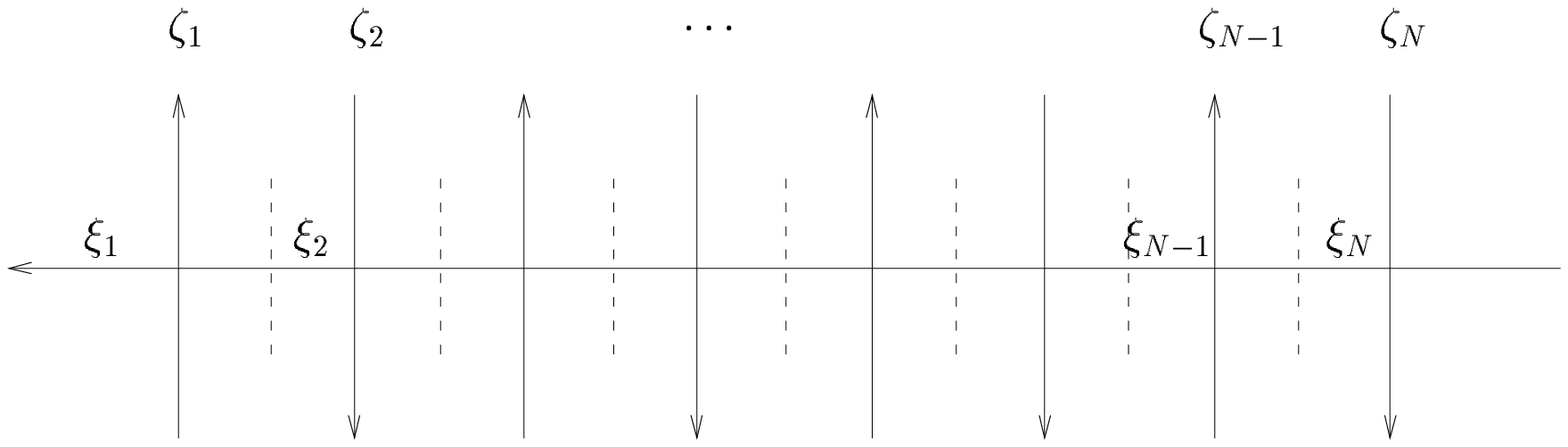}\vspace*{-128mm}%
\label{A1}\caption{The finite completely inhomogeneous transfer matrix
$T(\{\z_i\},\{\et_i\})$}\efig
\noindent It is the trace of the product of $N$ R-matrices,
each with independent vertical and horizontal rapidities
$\z_i$ and $\et_i$, $i=1,\cdots,N$. Two such transfer matrices
$T(\{\z_i\},\{\et_i\})$ and $T(\{\zp_i\},\{\etp_i\})$ commute
if $(\et_i/\z_i)/(\etp_i/\zp_i)=$const, independent of $i$.
This means that the normalised eigenvectors of
$T(\{\z_i\},\{\et_i\})$ depend only on the ratios
$(\et_i/\z_i)/(\et_j/\z_j)$ for $i \ne j$ \cite{Bax82}.

The first step is to specialise the transfer matrix to the
case when all the  horizontal rapidities on the left half of
the lattice are equal to  $\etp$, the left
vertical rapidities are alternately $\zp$ and $1/\zp$,
the right horizontal rapidities are all $\et$ and the right vertical
rapidities are alternately $\z$ and $1/\z$. Then we
identify the infinite product of such
transfer matrices (taking $N\ra \infty$ as well) with the
product $A_{SW}(\zp,\etp)A_{NW}(\z,\et)$ of suitably
normalised CTM's.
Such an identification implies
\be A_{SW}(\zp,\etp) A_{NW}(\z,\et) = \tau(\zp,\z,\etp,\et)
X(\zp,\z,\et/\etp) .\label{aax}\ee
Here and elsewhere we adopt the convention that lower case
letters represent scalar functions and that upper case letters
represent matrix functions.
We also suppress the $i$ superscripts on the CTM's.
Now send $\et \ra \c/\etp$ and
$\etp \ra \c/\et$ (where $\c$ is an arbitrary constant) and eliminate $X$.
Then,
\be \tau(\zp,\z,\chi/\et,\c/\etp) A_{SW}(\zp,\etp)A_{NW}(\z,\et) =
\tau(\zp,\z,\etp,\et) A_{SW}(\zp,\c/\et) A_{NW}(\z,\c/\etp)
\label{elim}. \ee
This equation immediately tells us that the product
\be \Ab_{NW}(\z,\et)= A_{NW}(\z,\et) A_{NW}(\z,\mu)^{-1}\label{redef1}\ee
(where $\mu$ is another constant) depends on $\z$
only through a scalar function.
Setting $\et=\mu$ in \rf{elim}, solving for $A_{SW}(\zp,\etp)$
and substituting in \rf{aax} gives
\be \Ab_{NW}(\z,\c/\etp) \Ab_{NW}(\z,\et) =
 c(\zp,\z,\etp,\et) Y(\zp,\z,\et/\etp), \label{baxprop}\ee
where
\bac
c(\zp,\z,\etp,\etp) &=& { \tau(\zp,\z,\etp,\et)
 \tau(\zp,\z,\c/\mu,\c/\etp)
\ov \tau(\zp,\z,\etp,\mu) },\label{redef1b}\\
Y(\zp,\z,\et/\etp) &=& A_{SW}^{-1} (\zp,\c/\mu) X(\zp,\z,\et/\etp)
 A_{NW}^{-1}(\z,\mu). \ea
(We shall suppress dependencies on the constants $\c$ and
$\mu$.)
Setting $\z=\zp=1$, we can rewrite \rf{baxprop} as
\be \Ab_{NW}(1,\etp) \Ab_{NW}(1,\et) =
d(\etp,\et) Z(\etp\et), \label{bax}\ee
where $d(\etp,\et)=c(1,1,\c/\etp,\et)$
and $Z(\etp\et)=Y(1,1,\etp\et/\c)$. We shall solve this equation
for $\Ab_{NW}(1,\et)$ and then reclaim
\be\Ab_{NW}(\z,\et)=r(\z,\et) \Ab_{NW}(1,\et),\label{redef2}\ee where
$r(\z,\et)$ is the scalar function
$r(\z,\et)=\Ab_{NW}(\z,\et)\Ab_{NW}^{-1}(1,\et)$.

Equation \rf{bax} is the generalisation of (13.5.17) of
Baxter's book \cite{Bax82}, and the argument now proceeds
as in \cite{Bax82} with some minor modifications.
Interchanging $\etp$ and $\et$ and eliminating $Z$ gives,
\be d(\et,\etp)\Ab_{NW}(1,\etp) \Ab_{NW}(1,\et) =
d(\etp,\et) \Ab_{NW}(1,\et) \Ab_{NW}(1,\etp). \label{comm1}\ee
Now consider a representation in which $\Ab_{NW}(1,\etp)$ is
diagonal. Then \rf{comm1}
implies
\bac d(\et,\etp)&=& d(\etp,\et),\\
\Ab_{NW}(1,\etp) \Ab_{NW}(1,\et) &=& \Ab_{NW}(1,\et) \Ab_{NW}(1,\etp).
\label{comm2}\ea
Thus $\Ab_{NW}(1,\etp)$, $ \Ab_{NW}(1,\et)$ and $Z(\etp\et)$ commute and have
common eigenvectors, independent of $\etp$ and $\et$.
We can clearly renormalise in the following way:
\bac \Ab_{NW}^R(1,\et) &=&
\Ab_{NW}(1,\et)/\a(\et),\\
Z^R(\et)&=& Z(\et)/\beta(\et), \label{redef3}\ea
where $\a(\et)$
and $\beta(\et)$ are the respective eigenvalues of some common eigenvector of
$\Ab_{NW}(1,\et)$ and $Z(\et)$. Then equation \rf{bax} becomes
\be  \Ab_{NW}^R (\etp) \Ab_{NW}^R(\et) =Z^R(\etp \et) \label{bax2}.\ee
Now we can diagonalise,
\be \Ab_{NW}^d(\et)=P^{-1} \Ab_{NW}^R(\et) P
\label{redef4}, \ee
where $P$ is the matrix of eigenvectors of $\Ab_{NW}^R(\et)$.
Define $Z^d(\etp\et)$
similarly such that
\be \Ab_{NW}^d(\etp) \Ab_{NW}^d(\et) = Z^d(\etp\et). \ee
Differentiating with respect to $\et$, it is apparent that the general solution
is
\be\Ab_{NW\,rr}^d(\et)=m_r\et^{-\a_r}\ws r=1,2,\cdots, \ee
where $m_r$ and $\a_r$ are independent of $\et$ and
$\z$. Using the redefinitions \rf{redef1}, \rf{redef2}, \rf{redef3} and
\rf{redef4}, we find,
\be A_{NW}(\z,\et) = a(\z,\et)  P\Ab_{NW}^d(\et) Q^{-1}(\z),\label{qeqn}\ee
where $Q^{-1}(\z) = P^{-1} A_{NW}(\z,\mu) $ and the scalar function
$a(\z,\et)=r(\z,\et)\a(\et)$.
Setting $\et=1$ in \rf{qeqn}, solving for $Q^{-1}(\z)$ and
substituting back into the same equation gives
\be A_{NW}(\z,\et) ={ a(\z,\et) \ov a(\z,1)} P\Ab_{NW}^d(\et)
\Ab_{NW}^d(1)^{-1} P^{-1} A_{NW}(\z,1).\ee
Defining the operator $D^d$ as that with diagonal entries $\a_1,\a_2,\cdots$,
gives
\be A_{NW}(\z,\et) ={ a(\z,\et) \ov a(\z,1)}\et^{-D} A_{NW}(\z,1), \ee
where $\et^{-D}=  P \et^{-D^d}P^{-1}$.

We may now specialise this formula to the case when $\z=1$ and
all the horizontal arrows of $A_{NW}$ point to the left
(the argument is
independent of the direction of the arrows). The CTM is that
of Figure 4, and is denoted
by $A_{NW}(\et)$. Since
$A_{NW}(1)=\id$, this gives
\be A_{NW}(\et)={\bar{a}(1,\et) \ov \bar{a}(1,1)}\et^{-D},\ee
where $D$ is unchanged and the bar indicates that the
scalar factors are those relevant to the homogeneous CTM
$A_{NW}(\et)$.
Hence we obtain the desired result
\rf{key} that
\bac A_{NW}(\et) A_{NW}(\z,\etp) &=& g(\z,\et,\etp) A_{NW} (\z,\et \etp)
\ws \hbox{where},\\
g(\z,\et,\etp) &=& {
\displaystyle{ a(\z,\etp) \bar{a}(1,\et)}
 \ov
\displaystyle{a(\z,\et \etp) \bar{a}(1,1)}}. \ea
\baselineskip=14pt
\pagebreak

\end{document}